# Variation in chemical composition and sources of PM$_{2.5}$ during the COVID-19 lockdown in Delhi

Chirag Manchanda [a], Mayank Kumar [a,*], Vikram Singh [b,*], Mohd Faisal [b], Naba Hazarika [c], Ashutosh Shukla [d], Vipul Lalchandani [d], Vikas Goel [a], Navaneeth Thamban [d], Dilip Ganguly [e], Sachchida Nand Tripathi [d,*]

a. Department of Mechanical Engineering, Indian Institute of Technology Delhi, New Delhi, India

b. Department of Chemical Engineering, Indian Institute of Technology Delhi, New Delhi, India

c. Department of Applied Mechanics, Indian Institute of Technology Delhi, New Delhi, India

d. Department of Civil Engineering, Indian Institute of Technology Kanpur, Uttar Pradesh, India

e. Centre for Atmospheric Sciences, Indian Institute of Technology Delhi, New Delhi, India

**Abstract**

The Government of India (GOI) announced a nationwide lockdown starting 25[th] March 2020 to contain the spread of COVID-19, leading to an unprecedented decline in anthropogenic activities and, in turn, improvements in ambient air quality. This is the first study to focus on highly time-resolved chemical speciation and source apportionment of PM$_{2.5}$ to assess the impact of the lockdown and subsequent relaxations on the sources of ambient PM$_{2.5}$ in Delhi, India. The elemental, organic, and black carbon fractions of PM$_{2.5}$ were measured at the IIT Delhi campus from February 2020 to May 2020. We report source apportionment results using positive matrix factorization (PMF) of organic and elemental fractions of PM$_{2.5}$ during the different phases of the lockdown. The resolved sources such as vehicular emissions, domestic coal combustion, and semi-volatile oxygenated organic aerosol (SVOOA) were found to decrease by 96%, 95%, and 86%, respectively, during lockdown phase-1 as compared to pre-lockdown. An unforeseen rise in O$_3$ concentrations with declining NO$_x$ levels was observed, similar to other parts of the globe, leading to the low-volatility oxygenated organic aerosols (LVOOA) increasing to almost double the pre-lockdown concentrations during the last phase of the lockdown. The effect of the lockdown was found to be less pronounced on other resolved sources like secondary chloride, power plants, dust-related, hydrocarbon-like organic aerosols (HOA), and biomass burning related emissions, which were also swayed by the changing meteorological conditions

[*] Corresponding Authors
 E-mail addresses : kmayank@mech.iitd.ac.in (M. Kumar); vs225@chemical@iitd.ac.in (V. Singh); snt@iitk.ac.in (S.N. Tripathi)

during the four lockdown phases. The results presented in this study provide a basis for future emission control strategies, quantifying the extent to which constraining certain anthropogenic activities can ameliorate the ambient air. These results have direct relevance to not only Delhi but the entire Indo-Gangetic plain (IGP), citing similar geographical and meteorological conditions common to the region along with overlapping regional emission sources.

**Summary of main findings**

We identify sources like vehicular emissions, domestic coal combustion, and semi-volatile oxygenated organic aerosol (SVOOA) to be severely impacted by the lockdown, whereas ozone levels and, in turn, low-volatility oxygenated organic aerosols (LVOOA) rise by more than 95% compared to the pre-lockdown concentrations during the last phase of the lockdown. However, other sources resolved in this study, like secondary chloride, power plants, dust-related, hydrocarbon-like organic aerosols (HOA), and biomass burning related emissions, were mainly driven by the changes in the meteorological conditions rather than the lockdown.

**Introduction**

The progression of air pollutants from their source to the receptor is governed by a multitude of transport processes (advective winds, convective updraft, or turbulent diffusion) and multiphase transformations. These transformations are often effected by chemical reactions that lead to heterogeneous mass transfer, which further complicates the system. Thus the concentration of each pollutant is often dependent on the concentration of other pollutants through a series of chemical reactions (Seinfeld, 2004). Time-resolved measurements of the concentration and chemical composition of these aerosols provide valuable insights into the levels of air pollution (Fehsenfeld, 2004). However, measurements alone are limited in space and time and are unable to provide much information on the origin or source of these pollutants. This has led to the widespread use of various data analysis/source apportionment techniques in the last few decades to extract more information on the nature and composition of emission sources (Watson and Chow, 2015). However, secondary pollutants display a highly non-linear relationship with precursor emissions; thus, source apportionment techniques are unable to provide any definitive information to predict the effect of increase/decrease of precursor emissions on secondary pollutant concentrations (Burr and Zhang, 2011).

The COVID-19 induced lockdown around different parts of the globe resulted in an unprecedented impact on the environment, with a drastic reduction in primary emissions; thus, enabling us to evaluate the impact of reduced precursor concentrations on primary and secondary aerosols, allowing us to better understand the dominant formation mechanism for a particular secondary pollutant in a given setting. The first lockdown was enforced in various parts of the Hubei province in China from 23$^{rd}$ January 2020, followed by similar measures in other cities (Wu et al., 2020). The COVID-19 outbreak was declared to be a pandemic by the World Health Organization (WHO) on 11$^{th}$ March 2020 (Sohrabi et al., 2020), following which lockdown or similar restricted movement measures were implemented in almost every region across the globe, although with varying stringency (Oxford COVID-19 Government Response Tracker, 2020).

An early study by Bao and Zhang (2020) analyzed the impact of reduced human mobility due to the lockdown on ambient air quality in 44 cities in Northern China from January to March 2020. The study found the average AQI to decrease by 7.8%, while significant pollutants like $SO_2$, $PM_{2.5}$, $PM_{10}$, $NO_2$, and CO, decreased 6.76%, 5.93%, 13.66%, 24.67%, and 4.58%, respectively. A consequent study by Li et al. (2020) utilized the Particulate Source Apportionment Technology (PSAT) coupled with the Comprehensive Air Quality Model with extensions (CAMx) to quantify contributions from 8 different sources to total $PM_{2.5}$ variations over the Yangtze River Delta (YRD) region, with the study period spanning 1$^{st}$ January to 31$^{st}$ March 2020 and the most stringent lockdown lasting from 24$^{th}$ January to 25$^{th}$ February 2020. Significant reductions in industrial operations, vehicular kilometers traveled (VKT), construction, and other anthropogenic activities were observed, in turn, bringing about a 25.4% to 48.1% decrease in $PM_{2.5}$ concentrations at different sites over the YRD region. However, an average rebound of 20.5% was recorded for ozone concentration. This anomaly was attributed to the fact that a significant drop in the $NO_x$ concentration was observed (29.5% to 51.7%). At the same time, the reduction in VOC was not as intense as $NO_x$ leading to a drop in titration effect towards ozone (Seinfeld, J.H. and Pandis, 2006).

Such increase in regional oxidation capacity effected by the rise in ozone concentration due to decrease in $NO_x$ in a VOC-limited environment during the lockdown was also reported by several other independent studies, such as Lv et al. (2020) for Beijing (China); Zheng et al. (2020) for Wuhan (China); Sicard et al. (2020) for Wuhan (China), Nice (France), Rome (Italy), Turin (Italy), Valencia (Spain); Tobías et al. (2020) for Barcelona (Spain); Mahato et al. (2020) for Delhi (India), Selvam et al. (2020) for Gujrat (India), Kumari and Toshniwal (2020) for Delhi (India) and Mumbai (India).

In India, the nationwide lockdown was implemented on 24$^{th}$ March 2020 and lasted up till 31$^{st}$ May 2020, with phase-wise relaxations starting 19$^{th}$ April 2020. Mahato et al. (2020) published an early work quantifying the impact of the first phase of the lockdown on ambient air quality in the Delhi-NCT region. The study reported average PM$_{2.5}$ and PM$_{10}$ concentrations to dip by 53% and 52% respectively when compared to the pre-lockdown, while SO$_2$, NO$_2$, CO, and NH$_3$ were found to decrease by 18%, 53%, 30%, and 12%, respectively. Similar trends in PM$_{2.5}$, SO$_2$, NO$_2$, and CO concentrations were observed independently by Srivastava et al. (2020) in both Lucknow and Delhi, by Kumari and Toshniwal (2020) in Delhi, Mumbai, and Singrauli, and by Selvam et al. (2020) in Gujrat. However, the reduction in SO$_2$ was found to be more pronounced in Mumbai (39%) Gujarat (48%) and was attributed to their closeness to the ocean and, thus, shipping emissions by Selvam et al. (2020) and Kumari and Toshniwal (2020).

Despite these early studies investigating the impact of the lockdown in India, there has been no study focusing on the variation of the sources and chemical composition of particulate matter during different phases of the lockdown or the impact of increased O$_3$ concentrations on the sources of PM$_{2.5}$. The present study is aimed at studying the highly time-resolved variation of sources contributing to both the organic and inorganic fragments of PM$_{2.5}$ along with black carbon, from pre-lockdown through each phase of the lockdown. These results aid us in quantifying the impact that reduction of certain primary emissions can have on overall air quality and the inadvertent effect these reductions can have on secondary aerosols. This study also extends the double positive matrix factorization (PMF) methodology proposed by Petit et al., (2014) to account for elemental, organic, and black carbon fractions of PM$_{2.5}$ in a single source apportionment analysis and utilize the elemental tracers to aid in understanding the source of organic aerosols.

## 2. Experimental Methods and Data Analysis

### 2.1 Sampling Site and Instrumentation

The sampling was conducted at the campus of Indian Institute of Technology (IIT), Delhi (28°32'N; 77°11'E). The instruments are housed in a temperature-controlled laboratory on the top floor of a four-story building on campus. The nearest source of local emissions is an arterial road outside campus, located about 150m from the building.

The nationwide lockdown was implemented in India for an initial period of 21 days, starting 25$^{th}$ March 2020 until 14$^{th}$ April 2020. The lockdown was extended for another 21 days until 3$^{rd}$ May 2020, with the first set of relaxations to certain agricultural and industrial activities, beginning 20$^{th}$ April 2020. Following the end of lockdown phase-2, the lockdown was further extended twice for a period of 14 days each, with increased allowances focused on restarting commercial activities before concluding on 31$^{st}$ May 2020. Further details about the allowances in each phase of the lockdown have been discussed in supplementary information (SI) section 3.

The study period has been subdivided into five phases, such that each subsequent phase coincides with increasing relaxations in the lockdown: Pre-Lockdown (PLD) (24$^{th}$ February – 24$^{th}$ March 2020), effective Lockdown Phase-1 (eLD-1) (25$^{th}$ March -19$^{th}$ April 2020), effective Lockdown Phase-2 (eLD-2) (20$^{th}$ April – 3$^{rd}$ May 2020), Lockdown Phase-3 (LD-3) (4$^{th}$ May – 17$^{th}$ May 2020), Lockdown Phase-4 (LD-4) (18$^{th}$ May – 31$^{st}$ May 2020).

The Xact 625i, XRF-based ambient metals monitor (Cooper Environmental Services, Tigard, Oregon, USA) equipped with a PM$_{2.5}$ inlet was deployed for sampling and was set up to quantify 36 elements with an hourly time resolution. Meanwhile, an Aerosol Chemical Speciation Monitor (ACSM, Aerodyne Inc., MA, USA) was deployed for analyzing the non-refractory particulate matter with an aerodynamic diameter smaller than 2.5µm (NR-PM$_{2.5}$). The ACSM measured the concentration of organic aerosols along with sulfate, nitrate, ammonium (SNA), and chloride concentration, with a time resolution of ten minutes, averaged to an hourly time scale. Black Carbon (BC) concentrations were measured using a multichannel Aethalometer (Magee Scientific Model AE33, Berkeley, CA) with a 2.5 µm inlet cut and a 1-minute time resolution.

Total PM$_{2.5}$ was measured using a collocated Beta Attenuation Monitor (BAM 1022, MetOne Instruments Inc., OR, USA), with a 15-min time resolution, due to technical difficulties faced during the lockdown, the data was only available for the PLD and LD-3 phase and is used for data quality assurance or quality control (QA/QC) purposes (Figure S1), Wind Speed (WS) and Wind Direction (WD) were calculated for the nearest grid point to the sampling site, using the Global Forecast System (GFS, NCEP, USA). Relative Humidity (RH), Ambient Temperature (AT), and rainfall were recorded using an onsite Ambient Weather Monitoring station. Ozone, SO$_2$, CO, and NO$_x$ measurements were taken from the Continuous Ambient Air Quality Monitoring Station (CAAQMS) at RK Puram, located at a distance of around 3 km from our sampling site. All CAAQMS stations in Delhi, including the RK Puram station, are managed by either the Central Pollution Control Board (CPCB) or the Delhi Pollution Control Council (DPCC) and together provide a network of near-real-time monitoring of PM$_{2.5}$ and PM$_{10}$ levels, along with major gaseous pollutants. Further details on the instrumentation and QA/QC checks for each instrument are provided in SI section S1.

**2.2 Source Apportionment using PMF**

This study utilizes Positive Matrix Factorization (PMF) (Paatero and Tapper, 1994) to apportion the measured particulate concentrations to realizable sources. PMF is a standard multivariate factor analysis tool widely used for source apportionment of aerosols (Sharma et al., 2016; Ulbrich et al., 2009; Vossler et al., 2016). The algorithm attempts to best describe the variability in a multivariate input dataset as the linear combination of a set of constant factor profiles and their relative contribution at every corresponding time step, as shown in Eq. (1):

$$x_{ij} = \sum_{k=1}^{p} g_{ik} f_{kj} + e_{ij} \quad \ldots\ldots (1)$$

Where $x_{ij}$ is the measured elemental concentration, $f_{kj}$ is the factor/source profile, $g_{ik}$ the time-varying contribution of each source, and $e_{ij}$ represent the elements of the residuals matrix. The indices $i$ and $j$ denote each of the n time steps and m chemical species, while $k$ refers to each factor/source out of total $p$ source profiles, which is defined by the user.

In PMF, each element of the factor matrix is constrained that no sample can have a negative factor contribution. The solution to Eqn. 1 is achieved iteratively by minimizing the object function or the goodness of fit parameter known as Q value:

$$Q = \sum_i \sum_j \left(\frac{e_{ij}}{s_{ij}}\right)^2 \quad \ldots.. (2)$$

Here, $s_{ij}$ corresponds to the measurement uncertainty for every cell of the input matrix $x_{ij}$. The PMF algorithm was solved using the Multilinear Engine (ME) -2 (Paatero, 1999). In this study, the PMF algorithm was implemented using EPA PMF 5.0 that is built on the ME-2 solution model. A detailed description of the model is provided in past studies (Paatero, 1997; Paatero and Tapper, 1994).

As discussed in section 1, the present study extends on the double PMF method, initially proposed by Petit et al. (2014). Petit and co-workers proposed to deconvolve the organic aerosol (OA) mass spectra using the routine PMF as described above, followed by a second PMF taking the deconvolved OA factors in conjunction with black carbon and inorganic ions like sulfate, nitrate, and ammonium as input. In the present study, we follow the same steps as proposed by Petit et al., (2014); however, instead of the inorganic ions, we use the elemental measurements made by the Xact 625i in conjunction with black carbon measurements and deconvolved OA mass spectra, for input to the PMF at the second step. Further details on PMF input preparation, factor selection, and uncertainty quantification for both the routine PMF and the double PMF have been reported in SI section S2.

## 3. Results and Discussion

The PLD phase in the present study marks the phase with no restricted movement, while eLD-1 corresponds to the phase with the most stringent lockdown. Consequently, the span of each phase from eLD-2 to LD-4 has been concomitant with increments in relaxations to the lockdown (resulting in increased commercial activity and human mobility) as implemented by the state/central government.

Variation in instrument total $PM_{2.5}$ and its constituents, along with the major gaseous pollutants for both during the lockdown in 2020 and the same period in 2019, is presented in Figure 1 (a-d). The instrument total $PM_{2.5}$ represents the sum total of the elemental, organic, and black carbon fractions of $PM_{2.5}$ measured using Xact 625i, q-ACSM, and Aethalometer AE33, respectively. The instrument total $PM_{2.5}$ includes the metals and chlorine measurements from the Xact, the organics and the inorganic ions (sulfate, nitrate, ammonium) from ACSM and BC from Aethalometer. The instrument total $PM_{2.5}$ has been compared with total $PM_{2.5}$ measured by a co-located BAM, and both have been found to have appreciable correlation (Pearson $R>0.91$) and low residual mass (less than 10%) (Figure S1). Considering the temporal variation of $PM_{2.5}$ during the lockdown phases (Figure 1(a)), we note that the average $PM_{2.5}$ values fall by 53.6% from PLD to eLD-1; this is in line with the findings of recent studies investigating the impact of the lockdown on $PM_{2.5}$ levels in Delhi (Dhaka et al., 2020; Kumar et al., 2020; Mahato et al., 2020; Yadav et al., 2020). The PM levels trend back towards the initial concentrations with increasing relaxations in subsequent phases of the lockdown; however, even during LD-4, average $PM_{2.5}$ remained 33% lower compared to the PLD values. Considering the $PM_{2.5}$ levels in 2019 (Figure 1(a)), we see that the average total $PM_{2.5}$ during the time period corresponding to eLD-2 to LD-4 (April to May 2019) is lower than what is observed during February to March 2019, even without the lockdown. This variation observed in 2019 potentially stems from meteorological parameters like boundary layer height, temperature, and RH, varying across seasons or due to the seasonal nature of some emission sources. Comparing the PM levels in 2020 with the corresponding periods in 2019; the average $PM_{2.5}$ levels were lower in 2020 by 27.8%, 53.1%, 36.8%, 34% and 18% for each of PLD, eLD-1, eLD-2, LD-3, and LD-4. These observations highlight the fact that any reduction during a lockdown phase in comparison with PLD cannot be attributed to as an impact of the lockdown alone, as that would undermine the inherent seasonal impact on the PM levels as observed for 2019. However, taking previous years' levels as reference for reduction during the lockdown cannot be justified either, as it can be seen that the PM levels in 2020 during PLD were ~28% lower compared to 2019; so while comparing with previous year levels may account for seasonal variation, it will discard any increase or decrease in source emission across the two years.

Taking note of the temporal variation of the gaseous pollutants, a significant drop in $NO_2$, NO, and CO concentrations 56%, 90%, and 32% (Figure S6(a)) respectively, was recorded during eLD-1 w.r.t PLD, while $SO_2$ remains largely unaffected by the lockdown. Comparing the levels of gaseous pollutants during the lockdown with the levels in 2019 (Figure 1(c) and Figure 1(d)), $SO_2$ in 2020 during the study period remains ~78% higher than the 2019 levels, reaching the highest in LD-4, where the average levels are around 2.8 times compared to the same period in 2019. NO2 and CO are on average 60.5% and 67% lower than their average levels in 2019, with the lowest during eLD-1with ~73% reduction as compared to 2019, highlighting the impact of the lockdown. NO concentration levels during the study period when compared to 2019, are severely impacted by the lockdown, while the average concentrations in 2020 are 46% less than 2019 during PLD, the average levels are 93 % lower in 2020 w.r.t 2019 during elD-1 to LD-4. Further discussion on time variations of these gaseous pollutants is presented in SI section 3.1.2.

The effect of the lockdown on secondary aerosol formation is further analyzed using the Sulfate Oxidation Ratio(SOR) and Nitrate Oxidation Ratio (NOR) (Figure 1(d)). The SOR and NOR are defined as the molar ratio of $SO_4$ and $NO_3$ to total oxidized S ($SO_4 + SO_2$) and total oxidized N ($NO_3 + NO_x$), respectively (Zhang et al., 2011). The average SOR for the PLD was 0.26, followed by 0.1, 0.15, 0.13, and 0.1 for each of eLD-1 to LD-4, while the average NOR for the PLD was 0.12, followed by 0.07, 0.1, 0.1, and 0.06 for each of eLD-1 to LD-4. According to previous studies, SOR and NOR lower than 0.25 and 0.10, respectively, are a marker for primary particulate matter (Ohta and Okita, 1990). Thus, the reduction in these ratios seems to indicate some role of lockdown in hindering the formation of secondary particles. However, contrary to the nature of total $PM_{2.5}$ and gaseous pollutants, these ratios do not tend towards the PLD values; with increasing relaxations, these ratios achieve the lowest average values in LD-4.

In Figure 1 (d), we also note that the behavior of $O_3$ doesn't reconcile with the trend followed by total $PM_{2.5}$ and other gaseous pollutants discussed above and are found to increase by 98%, 121%, 118%, and 54% in each of eLD-1 to LD-4 w.r.t the PLD concentrations (Figure S6(a)). Similar, anomalies in terms of increase in ozone concentration following the lockdown have been observed in some recent studies, not only in India but also in different parts of China, France, Italy, and Spain (Sicard et al., 2020; Tobías et al., 2020; Zheng et al., 2020) and have attributed the increase in ozone to the reduction of $NO_x$ in a VOC-limited environment (Monks et al., 2015). This non-linear coupling between VOC, $NO_x$, and $O_3$ was originally discussed in an early study by Finlayson-Pitts and Pitts (1993). However, it is also interesting to note that despite the increasing $NO_x$ concentrations in LD-4 and high $O_3$ concentrations during the same period, NOR remains 0.06, indicating low concentrations of particulate nitrate. The same scenario has been discussed by Finlayson-Pitts and Pitts (1993), presenting the hypothesis that VOC and $NO_x$ compete for OH radicals for oxidation. When the VOC to $NO_x$ ratio increases (decreasing $NO_x$ at constant VOC), the oxidation of VOC is favored over $NO_x$, resulting in lower nitrate concentrations. Similarly, for $SO_2$ and associated lower sulfate indicated by low SOR despite high $O_3$, gas-phase oxidation of $SO_2$, similar to $NO_2$, competes for the OH radical, while aqueous phase oxidation of $SO_2$ is limited by the acidity of the reaction products, as this oxidation route is efficient only near neutral conditions (Wilson et al., 1972). Further discussion on the variation of $PM_{2.5}$, its constituents, and gaseous pollutants is presented in SI section S3.1.

**3.1 Source Apportionment of elemental $PM_{2.5}$ (measured using Xact 625i)**

The elements measured using the Xact 625i were subjected to source apportionment using PMF. The input dataset was found to be best represented by a seven factor solution namely, vehicular emissions, biomass burning, secondary chloride, Zn-K-Br rich, dust related, power plants and local coal combustion. As discussed in supplementary information section S2.1, measurements corresponding to only 16 elements out of the 36 elements measured by the Xact, were utilized for input to the elemental source apportionment (SA). It is important to note while the considered elements along with the organic and black carbon measurements present closure to total PM2.5 (SI S1), the limited no. of elements may limit our understanding of some of the sources resolved by elemental SA. Sources like secondary chloride are dominated solely by chlorine, while in case of the Zn-K-Br rich source, none of the dominant species point towards a specific emissions source. A greater number of elemental measurements in the future may add more meaning to the sources resolved in this study, by providing additional tracers. Also, it could potentially aid in further resolving factors like the Zn-K-Br rich factor taking advantage of more marker species. The sources corresponding to the apportioned factors were assessed based on the species dominating every factor profile. Each of the species was quantified in two ways:

a) Based on the percentage of factor mass, given by the average concentration of the species of interest divided by the sum of the average concentration of each species within the factor

b) Based on percentage species across factors, given by the concentration of the species of interest in the factor under consideration divided by the sum of the concentration of the same species across all factors.

A detailed description of the resolved seven-factor solution (Figure 2) is as follows:

*3.1.1 Vehicular Emissions*

The vehicular emissions factor (Figure 2(a)) was found to be dominated by S (36%) in terms of the % factor mass, followed by Cl (19%), K (17%), Fe (13%), Ca (7%) and Zn (6%) respectively in terms of the percentage factor mass. However, in terms of the percentage species across factors, vehicular emissions accounted for 60% of the total Mn content, followed by 33% of Ba, 30% of V, 22% of Zn, 15% of Ca, 13% of K, and 11% of S.

Sulfur and Vanadium is known to occur naturally in crude oil; while pollution control measures have remained focused on reducing sulfur content in fuels, studies have pointed out the use of sulfur in engine oil anti-wear additives (Fitch, 2019). Multiple studies in the past have attributed Mn, Fe, Zn, and Ba to vehicular emissions, brake wear, and engine wear, in particular, recognizing them as abundant trace elements in brake pads and brake lining (Gianini et al., 2012; Grigoratos and Martini, 2015; Rai et al., 2020b; Thorpe and Harrison, 2008). Ti and V have also been attributed to brake and tire wear in some studies in the past (Gerlofs-Nijland et al., 2019). Potassium is noted to be used as an anti-freeze inhibitor and as an additive in engine oils. Also, K is known to be present in all unleaded fuels (Spencer et al., 2006). Calcium and Chlorine are known to be added to engine lubricants, Ca-compounds serves as a base to neutralize acids, while Cl-based additives act as dispersants to retain dirt in suspension, to protect the engine (Dyke et al., 2007; Lyyränen et al., 1999; Rudnick, 2017).

In terms of the time variation (Figure S3(a)), this factor is significantly affected by the lockdown with a 96% reduction in average concentration from PLD to eLD-1 (Figure S6(b)), the time series, and the composition pie-chart (Figure 2(b)) show a steady rise in the concentration of this factor, while the factor concentration in phase-4 remains 70% lower than its pre-lockdown value.

As an additional proxy, mobility trends (Google LLC, 2020) (Figure 2(c)) quantifying the percentage change in transit station mobility w.r.t PLD, was compared with the time variation of this factor, and a significant correlation (Pearson R = 0.81) between the two was noted.

In addition to the characteristic species noted above, this factor displayed a sharp diurnal peak coinciding with the morning rush hour and evening rush hour (Figure S4) during PLD and LD-4, when there was comparatively normal traffic load. During eLD-1 to LD-3, the vehicular movement has remained extremely restricted; thus, no diurnality in traffic-related emissions was found.

*3.1.2 Biomass Burning*

In terms of the % factor mass, the biomass burning factor (Figure 2(a)) was dominated by K (36%), followed by S (27%), Cl (21%), and Si (12%), respectively. Looking into the percentage species across factors, biomass burning was found accountable for 75% of the total Se content, followed by 58% of K, 19% of Si, and 14% of S.

Multiple studies across the globe have proposed the use of potassium as an elemental marker to identify biomass source profiles (Khare and Baruah, 2010; Pant and Harrison, 2012; Reche et al., 2012; Shridhar et al., 2010). Past studies have also reported Se to reach significant levels in biomass grown in selenium-rich soils (Goldstein, 2018), which in turn are common in northern India (Sharma et al., 2009). Silicon is known to be emitted from the pyrolysis of fibers in biomass like straw, cereal, and grass (Obernberger et al., 2006). Li et al. (2003) concluded that for fresh biomass burning plumes, most potassium exists as KCl, while in aged plumes, the chlorides are partly replaced by sulfates—thus providing evidence for high sulfur and chloride concentrations in conjunction with potassium in biomass associated source inventories.

The average factor concentration is found to lower by 25% in eLD-1 (Figure S6(b)); however, that may be a direct consequence of lower biomass burning emissions. Again it is not instinctive to assume any strong dependence of the lockdown event over biomass burning emissions. Thus it is difficult to ascertain whether the decrease in the concentration stemmed from the lockdown. The concentration starts increasing in April and peaks in May, which also coincides with the wheat harvesting season, thus resulting in increased residual crop burning activities (Jethva et al., 2019). The factor time series was also compared with satellite-based fire counts (LANCE FIRMS, 2020) in a 200 km radius of the sampling site (Figure 2(d)), and a significant correlation (Pearson R = 0.71) was observed between them.

*3.1.3 Secondary Chloride*

The Secondary Chloride factor (Figure 2(a)) is solely dominated by chlorine in terms of factor mass, with Cl accounting for 98% of the factor mass. Also, in terms of % species across factors, secondary chloride contributes to 89% of total measured Cl and 27% of total Br.

The diurnal variation associated with this factor (Figure S4) is very similar to that of $NH_4$, with a sharp peak around 6:00 IST. This leads us to the possibility of this factor stemming from ammonium chloride. With lower saturation vapor pressure, ambient ammonia and HCl may condense in the particulate phase and vaporize back with an increase in saturation vapor pressure after sunrise.

A similar factor from source apportionment of $PM_1$ in Delhi was observed by Jaiprakash et al. (2017). It was suggested that HCl fumes transported from metal processing plants to the north-west of Delhi reacted with the high ammonia concentrations in Delhi to condense in the particulate phase (Warner et al., 2017). Similar conclusions were reached upon by Gani et al. (2019) for the high particulate chloride concentrations observed by them.

The secondary chloride factor seems to have no dependence on the lockdown event as the average concentration of this factor increases marginally from PLD to eLD-1 (3.94 $\mu g/m^3$ to 4 $\mu g/m^3$) (Figure 2(b)).

From Figure S3(a), secondary chloride time series, we can note that almost all peaks for this factor correspond to a local wind direction in the sector of 302-333 degrees or NW direction. Also, the factor concertation is found to decrease noticeably in eLD-2 (1.25 $\mu g/m^3$) and remain low in LD-3 (1.26 $\mu g/m^3$) (Figure 2(b)). This again seems to have a dependence on the wind direction, starting from the beginning of eLD-2 up till the middle of LD-3, a shift in wind direction towards the southeast can be noted. However, the wind direction again shifts to the northwest towards the end of phase-3, which marks the rise in the secondary chloride concentrations again. In addition to the dependence on wind direction the reduction in average particulate-bound chlorine concentrations from PLD to LD-3 (3.94 to 1.26 $\mu g/m^3$), is also influenced by the rising ambient temperatures during the summer months (Figure S2), this is in line with the observations of past studies assessing the chloride levels in Delhi.(Gani et al., 2019; Rai et al., 2020a).

It is important to take note that the diurnal behavior associated with this source (Figure S4), reassures the existence of chlorine in form of $NH_4Cl$. Also, the dependence of this factor on the north-westerly winds as observed in this study, lend further support to the potential role of HCl fumes emitted in the north-west, neutralized by the high $NH_3$ levels in Delhi, as proposed by Jaiprakash et al., (2017). The possible sources of Cl or Br emissions may include a variety of sources like waste burning or industries; however, to account for the lack of dependence of this factor on the lockdown and following relaxations, the source of HCl/HBr must also remain unaffected/minimally affected by the lockdown.

*3.1.4 Zn-K-Br Rich*

In terms of % factor mass, this factor is mainly composed of Zn (42%), K (40%), and Fe (11%). However, in terms of % species across factors, these factors contribute to 68% of total Zn, 38% of total Br, and 16% of K (Figure 2(a)).

Multiple studies in the past have attributed a Zn-dominated factor to waste incineration (Gupta et al., 2012; Julander et al., 2014; Parekh et al., 1967; Sweet et al., 1993). These studies have mainly been associated with electronic or municipal waste burning, where a halide catalyzes the volatilization of metals to form metal halogenides, usual metals related to waste incineration in addition to Zn, include K, As, Fe, and Pb. While Cl is a more abundant halide, but Vehlow et al. (2003) have discussed how Br may be high in plastics containing flame retardants and, in turn, drive the volatilization of heavy metals like Zn Fe and As. However, Zn and As have also been attributed to iron/steel industries and waste incineration (Duan and Tan, 2013). Also, past studies have attributed Zn-Pb-Cl to industrial emissions (Bullock and Gregory, 1991).

This ambiguity in the published literature regarding tracer for waste incineration/industrial activities has lead us to define this factor as a Zn-K-Br rich factor only. In terms of the time variation (Figure S3(a)) the factor in line with total PM$_{2.5}$ decreases by 42% in eLD-1 w.r.t pre-lockdown, followed by 85%, 54%, and 50% in each of eLD-2 to LD-4 w.r.t PLD concentrations (Figure S6(b)).

*3.1.5 Dust Related*

The predicted dust-related source profile (Figure 2(a)) is dominated by Si, accounting for 48% of the factor mass, followed by 22% and 20% of the factor mass for Ca and Fe, respectively. In terms of % species across factors, dust-related source accounts for 84% of total Sr, 80% of total Si, 76% of Ca, 68% each of Ti and V, 55% of Fe, and 37% of Mn. Each of the above-noted species has been extensively used as tracers for road dust/crustal elements in multiple studies across the globe (Gupta et al., 2007; Kothai et al., 2011; Rai et al., 2020a; Sharma et al., 2016; Sun et al., 2019).

In terms of the time variation (Figure S3(a)) of this factor, there seems to be no observable effect of the lockdown on dust-related particulate matter. However, we observe a significant correlation of the factor concentration with ambient temperature (Pearson R= 0.64) (Figure S2(a)) and an inverse correlation with RH (Pearson R= -0.67) (Figure S2(a)), which is in agreement with several past studies (Csavina et al., 2014; Jayamurugan et al., 2013). Also, during LD-4, we observe an increase in the average concentration of this factor, from 1.15 $\mu$g/m$^3$ in LD-3 to 2.77 $\mu$g/m$^3$ in LD-4 (Figure 2(b)), which may be influenced by multiple meteorological parameters like WD, WS or gust events.

*3.1.6 Power Plants*

Considering the % factor mass, sulfur solely dominates this factor profile accounting for 93% of the total factor mass. In terms of the % species across the factors, the power plants factor contributes to 73% of the total sulfur, 35 % of As, and 32% of total Ba (Figure 2(a)).

In this study, the power plants factor (Figure S3(a)) displays a significant correlation with the SOR (Figure 1(d)) (Pearson R = 0.922), signaling towards the sulfur content in the particulate phase is actually in the form of aqueous sulfate. Also, past studies evaluating power plant emissions as well as source apportionment studies have highlighted the use of As and Ba as tracers for coal-based power plant emissions (Reddy et al., 2005; Zhao et al., 2017; Zoller et al., 1974).

Observing the time series of this factor, we do note a 65% decrease in eLD-1 w.r.t the PLD concentrations (Figure S6(b)). A recent report from the Power System Operation Corporation (POSOCO, 2020) does indicate a significant reduction (44% reduction in April compared to last year) in the power demand due to the closure or scaled-down operations in almost all industries due to the lockdown, which in turn could lead to temporarily scaled down operations at some power plants. Thus, some order of reduction in source emission can also be partially responsible for the significant drop observed in the factor concentration at the receptor.

However, it would be implausible to attribute the entire reduction to the lockdown alone, as there has been significant variation in the factor concentration within the PLD (-54% to +89% w.r.t PLD average) (Figure S6(b)), indicating some role of meteorological or other transport variables rather than the source emission alone for the variation in the concentration. Again, during eLD-2, the average concentration is found to increase relative to PLD levels; however, the concentration again starts to fall during LD-3 and LD-4, thus advising of some external metrological/transport phenomena affecting the concentration values.

*3.1.7 Local Coal Combustion*

The coal combustion factor (Figure 2(a)) is dominated by Lead, Zinc, and Sulfur accounting for 66%, 21%, and 8.5% of the total factor mass, respectively. Considering the % Species across factors, coal combustion is responsible for 87% of Pb, 18% of Se, 16% of As, and 9% of Zn. While coal combustion is found to account for only 0.3% of total sulfur, it is essential to note that the % contribution of this factor to elemental $PM_{2.5}$ has remained quite low (less than 1.9%) throughout the study period.

As and Se have been widely used as markers for coal combustion (Gupta et al., 2007; Hien et al., 2001; Lee et al., 2008; Sharma et al., 2007). Zn again has been used as a marker for coal combustion in India due to relatively higher Zn content in Indian coals (Almeida et al., 2006). While commercially available coal has lower Pb content, Negi et al. (1967) reported the higher concentrations of Pb and Zn in domestic soft coal. It is also important to note that domestic Indian coals have been found to have low sulfur content (less than 0.6% by mass), with an exception to coal deposits in north-eastern India with high sulfur content (Chandra and Chandra, 2004; Sarkar, 2009). Also, past studies have reported Indian power plants to utilize blends of imported and domestic coals supporting the higher sulfate emissions from power plants (Central Electricity Authority, 2012; Chandra and Chandra, 2004).

Evaluating the temporal variation associated with this factor (Figure S3(a)), we notice that the lockdown implementation brings about a 95% reduction in the average concentration of the coal combustion source, comparing eLD-1 to PLD conditions. With increasing relaxations, the percentage reduction in average concentration w.r.t the PLD falls to 90% in eLD-2 and LD-3 and finally 85% in LD-4 (Figure S6(b)).

Since the source profiles and supporting literature indicate domestic soft coal burning, the real-world sources may be connected to small scale industrial setups, eateries, or household usage of domestic grade coal, and such sources appear to be drastically affected by the lockdown and display only a marginal increase in emissions even with increasing relaxations. Such a variation could possibly stem from the massive outflow of migrant laborers from the NCT region, resulting in the sudden downfall of domestic coal usage for cooking purposes (Roy and Agarwal, 2020).

**3.2 Source Apportionment of Organic Aerosols (measured using Q-ACSM)**

The organic content of the total $PM_{2.5}$ mass is subjected to source apportionment using PMF. A six-factor solution was found to fit the input data best. The apportioned factors were identified by the mass spectra signatures, their correlation with tracers, and their diurnal behavior (Ulbrich et al., 2009; Zhang et al., 2005b). The present study further correlates each apportioned factor to corresponding reference factor profiles from Ng et al. (2011). Figure 3 presents the predicted mass spectra for each profile, along with their temporal variation and correlation with external markers. The detailed description of each predicted source profile is as follows:

*3.2.1 Semi-Volatile Oxygenated Organic Aerosol (SVOOA)*

The factor profile, as seen in Figure 3(a), is characterized by a significant peak at *m/z 43*, which is a characteristic of less oxidized secondary organic aerosol (Li et al., 2019; Zhu et al., 2018). The diurnal variation (Figure S5) of this factor presents two peaks: early morning (6:00 – 9:00 IST) and a weaker peak at midnight, signaling SVOOA concentrations to be affected by photo-oxidation of fresh emissions (morning peak) along with boundary layer height (midnight peak), similar to observations made by Chakraborty et al. (2018). The SVOOA source profile predicted from the PMF analysis was noted to have a Pearson R correlation of 0.93 with the reference SVOOA profile from Ng et al. (2011)

The SVOOA factor time series displayed a strong correlation with $NO_3$ (measured using ACSM) (Pearson R = 0.95) (Figure 3(c)), which is in agreement with the trend reported in past studies (DeCarlo et al., 2010; Dzepina et al., 2009; Volkamer et al., 2006), and is attributed to the analogous semi-volatility of SVOOA and nitrate resulting in similar gas-particle partitioning.

The SVOOA factor drops significantly (86%) after the lockdown is implemented; the emissions increase with increased relaxations from eLD-2 to LD-3; however, there is a small drop (18%) in average concentration again from LD-3 to LD-4 (Figure S6(c)).

*3.2.2 Hydrocarbon-like Organic Aerosol (HOA)*

Alkyl fragment signatures distinctly mark this factor profile (Figure 3(a)) with prominent contributions from *m/z 55* and *57* (Aiken et al., 2009; Ng et al., 2011). The resultant HOA profile has a strong correlation (Pearson R = 0.95) with reference HOA spectra from Ng et al. (2011).

Past studies have found HOA to correlate well with black carbon (BC) (Mohr et al., 2009; Sun et al., 2016). In the present study, we note an excellent correlation between BC and HOA (Pearson R = 0.96) (Figure 3(c)) during the PLD phase; however, post-lockdown, the trend of BC and HOA become completely disparate, resulting in a negligible correlation between the two. The significant correlation with BC often is taken as support for the vehicular origin of HOA (DeWitt et al., 2015).

However, from section 3.1.1, we note the vehicular emissions to drop significantly post-lockdown (96%) while HOA concentrations lower only by 14% post-lockdown. At the same time, it departs from the trend followed by BC (Figure 2(d)), indicating that during the lockdown, HOA originates from a source other than vehicular emissions. This uncharacteristic loss of correlation with BC and the potential sources supplementing HOA during the lockdown will be investigated further in section 3.3.

The initial studies that looked into the deconvolution of HOA from POA (Zhang et al., 2005a) suggested its connection to vehicular origin based on the significant correlation with vehicular markers like $NO_x$ and BC and the fine mode of particulate matter corresponding *m/z 55* and *57* as compared to *m/z 44* which grows larger while aging. Zhang et al. (2005a) also discussed car-chaser and lab-based studies, wherein both diesel emissions and lubricant combustion resulted in HOA like spectra, as heavy oils, lubricants, cooking oils are known to correspond to *m/z 55*, while mass spectra associated with gasoline and diesel-like fuels displayed a more definite *m/z 57* peaks.

Hao et al. (2014) also observed appreciable HOA contributions in a low-traffic village setting. They attributed the source to be a combination of industrial, cooking, and biomass burning along with the low contribution from traffic.

Thus, it may be hypothesized that either HOA during the lockdown originates from diesel/lubricant based emissions from sources other than vehicles, like diesel-based generators in industries or cooking-related activities. However, similar to SVOOA, even HOA experiences a sharp reduction towards the end of LD-4. The potential causes of this erratic behavior would be investigated further in section 3.3.

*3.2.3 Biomass Burning Organic Aerosol (BBOA-1 and BBOA-2)*

In the present study, we resolve two separate biomass burning related factors (Figure 3(a)). However, BBOA-1 can be categorized as a fresh/ primary emission, with its mass spectra highly correlated to the BBOA profile from Ng et al. (2011) (Pearson R = 0.937). For the other BBOA profile, i.e., BBOA-2, we observe enhanced concentrations of *m/z 43* and *m/z 44*, indicating that BBOA-2 is relatively aged. Also, the diurnal variation (Figure S5) associated with BBOA-1 displays primary emission like behavior, with early morning peaks, while BBOA-2 also displays peaks around noon, which is a characteristic of *m/z 44* or $CO_2^+$, indicating the possibility of regionally transported emissions responsible for BBOA-2.

BBOA-2 also shares a good correlation with the Ng et al. (2011) reference BBOA profile (Pearson R =0.82). Both BBOA-1 and BBOA-2 are marked by intensified peaks corresponding to *m/z 60*. Levoglucosan is known to be proportional to $C_2H_4O_2^+$ (a fragment at *m/z 60*) is extensively used as a marker for biomass burning in AMS-based studies (Aiken et al., 2009).

It is also important to note that while both the BBOA factors present a good correlation with the reference BBOA profiles, these factors profiles do share similarities with Coal Combustion Organic Aerosol (CCOA) detected by multiple studies in the past (Huang et al., 2014; Zhu et al., 2018). However, resolving the CCOA profile during ambient sampling is often contingent upon tracer fragments like PAHs and $C_6H_{10}O^+$ and heavier m/z's (Lin et al., 2017), which can be measured using the HR-ToF-AMS or ToF-ACSM. Since, the present study utilizes a q-ACSM limiting the m/z spectra to 120, the BBOA factors may have a trace or high contributions from CCOA factor that cannot be resolved with certainty with the single PMF due to lack of tracer ions. Past studies have also noted that splitting of factors, in absence of strong tracers, is most likely a mathematical artifact rather than a true component (Ulbrich et al., 2009). Nevertheless, the potential association of the BBOA factors with coal combustion or related activities, is further investigated in section 3.3 using double PMF and elemental tracers.

Similar to the biomass burning factor in section 3.1, it is not intuitive to presume an effect of the lockdown on biomass burning. However, comparing the two BBOA factors, we see that BBOA-1 displays higher concentrations in pre-lockdown, while reduced order (76%) values following lockdown steadily rise from eLD-1 to LD-3. BBOA-2 displays lower concentrations in the PLD period and steadily rise to a maxima (5.1$\mu$g/m$^3$) up till LD-3 (Figure 3(b)). Thus, indicating that while the lockdown leads to a decrease in the primary BBOA emissions (BBOA-1), it in some way enhances the regional transported/aged fraction of BBOA (BBOA-2). However, similar to both SVOOA and HOA, both BBOA factors significantly reduce both in absolute concentration and percentage contribution in LD-4 (Figure S3(b) & Figure 3(b)).

Interestingly none of the BBOA factors display any positive correlation or similar trend with the satellite-based fire counts (LANCE FIRMS, 2020) (Figure 3(c)), which was seen in the potassium dominated biomass burning factor in section 3.1 and neither with the biomass burning factor resolved from elemental SA.

A potential contributing factor to this discrepancy may stem from the stance that the tracers used to resolve these factors, i.e., *m/z 60* and K$^+$, point to different types of combustion processes. A study by Brown et al. (2016) presented a comparison between different biomass burning markers like K$^+$, BC, and levoglucosan or, in turn *m/z 60*. It was noted that K$^+$ and BC are more prominent products in flaming combustion (which is usually captured as fire counts). However, levoglucosan is a more prominent emission in the case of smoldering combustion (Lee et al., 2010).

Another probable cause may be that BBOA emissions can be associated with multiple sources and may not yield a good correlation with a single K$^+$ resolved biomass burning source alone, rather a combination of sources. This possibility is investigated further in section 3.3.

*3.2.4 Low Volatile Oxygenated Organic Aerosol (LVOOA-1 and LVOOA-2)*

LVOOA is addressed as an aged or oxidized aerosol and is majorly marked by a distinct peak of *m/z 44* or CO$_2^+$. In this study, we resolve two LVOOA factors (Figure 3(a)), i.e., LVOOA-1 and LVOOA-2. Both the LVOOA factors display a strong correlation with the Ng et al. (2011) reference spectra (Pearson R =0.95, 0.93 for LVOOA-1, and LVOOA-2, respectively).

Observing the temporal variation, we see that LVOOA-2 is at a high concentration (9.1 $\mu$g/m$^3$), which reduces after the lockdown (2$\mu$g/m$^3$) and steadily rises to a noticeably high concentration in LD-4 (12.3$\mu$g/m$^3$). On the other hand, LVOOA-1 mostly remains at a lower concentration from PLD to LD-3 ( from 2 $\mu$g/m$^3$ to 4 $\mu$g/m$^3$); however, it rises to a significant concentration in LD-4 (10.6 $\mu g/m^3$) (Figure 3(b)). It is important to note that all apportioned organic factors decrease considerably in LD-4, while both the LVOOA factors experience a significant rise.

LVOOA is known to correlate well with sulfate, citing the similar low-volatility observed in both species (Zhu et al., 2018). In Figure 3(c), we see that LVOOA-1 neither follows the trend nor is correlated to sulfate, whereas LVOOA-2 expresses a significant correlation with sulfate (measured using ACSM) during PLD (Pearson R = 0.93) and continues to display a significant correlation up till LD-2 (Pearson R =0.75); however, the correlation significantly deteriorates in LD-4. The decreased correlation with sulfate may point towards a change in source contributing to LVOOA-2 in LD-4, which will be investigated further in section 3.3.

Investigating the diurnal variations associated with both factors (Figure S5), we see that LVOOA-2 displays a flat diurnal profile with a marginal peak at noon, which is characteristic of $CO_2^+$ formation by photochemical oxidation. However, if we look at the phase-wise diurnal variation, the diurnal profile for LVOOA-1 in lockdown phase-4 behaves like a primary pollutant diurnal with an early morning peak rather than the afternoon peak, suggesting a primary aerosol-like formation mechanism for LVOOA-1.

Earlier in section 3, we noted low SOR and NOR values, which usually correspond to primary rather than secondary aerosols. We must also take into account that LD-4 began right after the period when ozone had reached its peak concentrations (section 3.1.2).

All these facts taken collectively indicate that in some way, all primary and intermediate organic aerosols are chemically aged in the presence of high $O_3$ to form LVOOA, leading to a reduction in all other organic factors and a significant rise in LVOOA. Jimenez et al. (2009), on a similar note, stated that the atmospheric oxidation of organic aerosol (OA) converges to LVOOA regardless of the original OA source. However, it is also important to note that the diurnal variation of LVOOA-1 hinted towards its primary emissions. Another study by Liggio and Li (2013) suggested a mechanism for the formation of oxygenated primary organic aerosols by uptake of primary oxygenated organic gases to aerosols, and thus presents a possible explanation for the primary rather than secondary origin of LVOOA-1.

**3.3 Combined source apportionment of elemental PM$_{2.5}$, black carbon, and organic PM$_{2.5}$**

In this section, we further resolve the sources of organic aerosols and interpret them better using elemental markers measured using the Xact 625i. We subject the elemental PM$_{2.5}$ and black carbon measurements from the Xact 625i and Aethalometer, respectively, along with the organic aerosol factors resolved in section 3.2, to source apportionment using PMF. As discussed in section 2.2 and in further detail in SI section S2.1.1, our efforts build upon the work of Petit et al., (2014), which introduced the double PMF methodology initially. While sections 3.1 and 3.2 have elucidated the impact of the lockdown on the elemental and organic fragments of $PM_{2.5}$ separately, this section gives a broader picture of the impact of the lockdown on sources encompassing both the organic and elemental fragments of $PM_{2.5}$ along with black carbon. This section also bridges the organic aerosol factors apportioned in section 3.2 to more realistic real-world sources, using elemental tracers. The implementation of the double-PMF is discussed in SI S2.2 and in greater detail in Petit et al., (2014)

The double-PMF analysis results in nine source profiles, i.e., vehicular emissions, biomass burning, secondary chloride, Zn-K-Br rich, dust-related, power plant, local coal combustion, combustibles factor (CF-1), and LVOOA dominated that is found to best explain the variability of the input dataset. The results of the double-PMF analysis are presented in Figure 4; the resolved source profiles as seen in Figure 4(a) highlight the fact that seven (vehicular emissions, biomass burning, secondary chloride, Zn-K-Br rich, dust-related, power plant, local coal combustion) out of the nine resolved factors are extremely similar to the 7-factor solution resolved from the elemental source apportionment in section 3.1. The elemental fraction of the double-PMF resolved sources are found to have an appreciable correlation (Pearson R > 0.9) with their counterparts in the elemental-PMF sources, and the same has been utilized to name the double PMF factors. The remaining two factors are dominated by the factors resolved in organic source apportionment instead of the elemental tracers and are named accordingly.

*3.3.1 Vehicular Emissions:* In addition to being marked by high concentrations of S, K, Fe, and tracers like Mn and Zn, as seen in the elemental SA factor, the double PMF apportions 35% HOA, 33% BC, and 18% SVOOA to vehicular emissions. As discussed in section 3.2.2, the temporal variation of HOA was well correlated with BC during PLD but deviated from the trend followed by BC post-lockdown. The temporal variation of vehicular emissions source (Figure 4(b)) and the fractional contribution of the vehicular emissions source to the BC and HOA time variation (Figure 4(c)) supports the vehicular origin of HOA and BC during PLD and LD-4, thus explaining the good correlation during those periods and also justifying the observed loss of correlation, from eLD-1 to LD-3 when the vehicular emissions fall down and no longer remain a major contributor to BC or HOA.

*3.3.2 Biomass Burning:* In line with its counterpart resolved through elemental source apportionment discussed in section 3.1.2, the double PMF resolved biomass burning source is marked by the high levels of K, S, Cl, and Se. Considering the organic SA factors, 40% of LVOOA-2, 35% of BBOA-2, 19% of BC, and 11% of SVOOA have been apportioned to the biomass burning source. As discussed in section 3.2.4, both apportioned LVOOA factors were found to increase in LD-4, which in turn also saw a rise in fire counts and biomass burning factor resolved by the elemental SA. A substantial part (40%) of LVOOA-2 being apportioned to double PMF biomass burning factor highlights the contribution of the biomass burning related activities in the rise of LVOOA-2 during LD-4. This impact can be clearly observed in Figure (4(c)), where biomass burning emissions are the major contributor towards LVOOA-2 during LD-3 and LD-4. It is also interesting to take note that while the BBOA-2 factor as a whole didn't present an appreciable correlation to elemental SA biomass burning factor, the double PMF connects the biomass burning activities marked by potassium tracers to contribute to a significant fraction, i.e., 35% of relatively aged BBOA or BBOA-2 on an average and can be clearly seen to dominate the BBOA-2 temporal variation during LD-3 and LD-4 (Figure 4(c)).

*Secondary Chloride* and the *Zn-K-Br Rich factor,* while are well correlated with their counterparts resolved with the elemental SA, both in terms of the source profile and temporal variation, double PMF attributes none of the organic emissions to these factors. This may be due to the unlikely event that these sources are not responsible for any organic emissions, or the tracers are impacted by some meteorological conditions or chemical transformations, rendering them unable to resolve the associated organic fragments. Thus, their source profile and temporal variation can be completely explained by their counterparts resolved by elemental SA discusses in section 3.1.3 and 3.1.4, respectively.

The *Dust Related factor,* similar to the secondary chloride and Zn-K-Br rich factors, doesn't contribute substantially to the organic fraction of $PM_{2.5}$. The double PMF does highlight some minor contributions (< 5% of species sum) of this source to LVOOA, BC, and BBOA-1; however, these contributions may result from contaminants common to dust-related sources, as discussed by, due to their interaction with multiple emission sources resulting in deposition of contaminants on dust-related PM.

*3.3.3 Power Plant*: As discussed for the elemental SA in section 3.1.6, in case of the double PMF resolved power plants factor too is marked by S, As, and Ba. However, it is interesting to note that that the power plant factor accounts for 56% of LVOOA-2, 54% of SVOOA, 38% of BBOA-1, and 13% of BC. As discussed in section 3.1.6, the sulfur content measured is mostly in the form of sulfate, which in turn is found to correlate well with LVOOA-2 up to LD-3, as discussed in section 3.2.4 (Figure 3(c)). The fractional contribution of this source to LVOOA-2 (Figure 4(c)) demonstrates that the power-plant emissions contribute significantly to total LVOOA-2 levels up to LD-2 (after which biomass burning dominates), justifying the appreciable correction of sulfate and LVOOA-2 up till LD-2 and its deterioration thereafter. It is also interesting to note that the power plant source, while being responsible for a significant amount of LVOOA-2, which is usually associated with regionally transported of aged OA, is also found to contribute to around 54% of SVOOA, which is more often linked with local, or moderately aged OA. The authors feel further investigation through back trajectory analysis may aid in understanding the origin of these source emissions to help understand the reason behind its significant contribution to both LVOOA-2 and SVOOA.

As discussed in section 3.2.3, the inability of the ACSM to measure higher m/z fragments, especially PAHs, renders the PMF solution unable to resolve the CCOA or COA with certainty. The power plant emissions contributing to the BBOA-1 levels (Figure 4(c)) indicates the possibility of BBOA-1 having a coal combustion related origin.

*3.3.4 Local Coal Combustion:* Similar to the elemental SA (section 3.1.7), the coupled SA coal combustion factor is dominated by Pb, accounting for more than 80% of total Pb. In terms of the organic content, the double PMF accounts for this factor to contribute to 52% of the total BBOA-1on average (Figure 4(a)). The presence of BBOA-1 in the power plants factor discussed earlier indicated the possibility of CCOA origin of BBOA-1; the presence of BBOA-1 in the coal combustion factor further reinforces this hypothesis that BBOA-1 has a coal combustion related origin. The same can be reaffirmed through the high fractional contribution of the local coal combustion factor to the BBOA-1 levels (Figure 4(c)). This observation also highlights another advantage of double PMF as being used to better resolved or identify ACSM based organic sources, even when other organic tracers are missing. Other than BBOA-1, 10% of total BC is also attributed to the local coal combustion source.

*3.3.5 Combustibles Factor-1:* As discussed earlier, unlike the other coupled SA sources, the CF-1 and LVOOA-dominated sources are marked by organic SA factors as tracers rather than the elemental markers. The CF-1 factor contributes to 65% of HOA and 64% BBOA-2 along with 19% of BC, and 12 % of S. The BC and sulfur content indicate some form of combustion, while both HOA and BBOA-2 signal towards the primary nature of the emissions associated with this source. The fractional contribution of the CF-1 source (Figure 4(c)) to various coupled factors indicates that the CF-1 source supplemented the HOA resolved from organic SA during the lockdown, which prevented HOA from reducing sharply with plummeting vehicular emissions. Based on the fractional contribution of these sources to BBOA-2 (Figure 4(c)), the CF-1 source dominates the BBOA-2 levels up till eLD-2. As discussed in sections 3.2.2 and 3.2.3, both HOA and BBOA-2 may be related to cooking activities, which implies that the CF-1 source emissions may stem from cooking activities, waste burning, or some other form of combustion activity, however, due to the lack of a reliable/ well-accepted marker to identify the actual source of this factor, we continue to refer to as an CF-1 source.

*3.3.6 LVOOA-dominated:* Like the CF-1 factor, this factor is populated by organic markers and accounts for 87% of total LVOOA and 13% of SVOOA. Interestingly, the factor is independent of any elemental/organic factor except LVOOA-1. As discussed in section 3.2.4, LVOOA-1 displayed diurnal behavior similar to primary organic aerosols, and now the LVOOA-1 factor not considerably associating with any other organic or elemental marker as resolved by double PMF further supports the hypothesis of the increased $O_3$ levels giving rise to increased LVOOA-1 levels in LD-4.

In addition to further apportioning the organics PMF derived factors, the elemental tracers, also aid in apportioning black carbon to real world sources, via the double PMF. As discussed earlier in account of the double-PMF resolved vehicular emissions, the notable contribution of the vehicular emissions to BC and HOA during PLD and LD-4, allows us to interpret the good correlation between HOA and BC, during these periods and loss of correlation during other times. A recent work by Goel et al., (2021), discussed the variation of black carbon sources during the lockdown, utilizing the wavelength-based two-component Aethalometer model proposed by Sandradewi et al., (2008). The original work proposing double-PMF (Petit et al., 2014), too utilizes the Aethalometer model, derived sources as input to the second PMF, in the two staged double PMF. However, in case of the present study we note that the BC sources' time variation derived using the Aethalometer model were highly correlated with each other (Pearson R >0.9). Since the PMF technique is limited in its ability to resolve highly correlated input data (Ulbrich et al., 2009), we use total BC as input to the double PMF rather than the Aethalometer model derived BC fractions. The authors believe the use of the elemental tracers in the double PMF methodology, allows for an independent method to apportion BC; however, the comparison of the results with those of the two-component model will prospectively be explored in future works.

## 4. Conclusions

The COVID-19 lockdown resulted in an unprecedented decline in anthropogenic activities, which in turn led to a considerable reduction (~ 54%) in ambient $PM_{2.5}$ levels. The detailed source apportionment results presented in this study reveal the varying impact of the lockdown on different sources contributing to the elemental and organic fractions of $PM_{2.5}$. Source apportionment of elemental $PM_{2.5}$ yielded seven source profiles; the vehicular emissions, coal combustion, and Zn-K-Br rich sources were severely impacted by the lockdown. However, the lockdown seemed to have minimal or no impact on biomass burning and dust-related sources. The dust-related factor displayed dependence on meteorological factors, while increased biomass burning emissions coincided with the crop burning season. The power plants-related elemental PM seems to be affected by both the lockdown and meteorological parameters. Interestingly, the secondary chloride factor observed elevated concentration peaks majorly from the north-west direction and remained largely unaffected by the lockdown.

The organics-only PMF resulted in 6 factors, i.e., SVOOA, HOA, two BBOA, and two LVOOA factors. The lockdown seems to have an appreciable effect on SVOOA factor concentrations with a reduction (86%) in eLD-1, followed by increased concentrations with relaxations in the lockdown. The fresh BBOA emissions (BBOA-1) decline following the lockdown, while the aged BBOA emissions(BBOA-2) rise, signaling intensified transport of BBOA related emissions from regional sources following the lockdown. HOA concentrations were marginally affected by the lockdown indicating sources other than vehicular emissions played a dominant role in HOA related emissions, contrary to the belief of HOA being dominated by vehicular emissions. The organic aerosol (OA) source apportionment also highlights a sharp rise in the LVOOA concentrations in LD-4 accompanied by a concomitant decay in concentrations of all other resolved OA sources; this rise is attributed to the oxidation of primary OA due to high ozone concentrations.

The double PMF implemented in the present study enabled much better interpretation of the temporal variation of the organic sources resolved by the organic SA, and successfully connected the elemental SA and organic SA results to give a complete picture of the impact of the lockdown on total $PM_{2.5}$ rather than just one the organic and inorganic fractions individually.

This is also the first study to quantify the impact of the COVID-induced lockdown on highly time-resolved sources of ambient $PM_{2.5}$ in India. These results have important implications for guiding future policies targeted and decreasing PM levels in not only Delhi but the entire IGP so that the actions are targeted on actual sources of emission, knowing the level of impact a particular source has on total PM levels. The results also highlight a prime concern for driving future emission control strategies, especially upcoming vehicular emission standards like Bharat Stage 6 (BS-VI), which may realize the low $NO_x$, VOC-limited setting without a lockdown, and lead to an inadvertent rise in ozone and LVOOA. The use of double PMF demonstrated in this study has clear implications to improve interpretation of the sources of organic aerosols and black carbon, which can aid future work in this domain.

**Acknowledgments**

The authors would like to acknowledge the IRD Grand Challenge Project grant, IIT Delhi (Grant No. IITD/IRD/MI01810G), Ministery of Human Resource Development (MHRD), Government of India and the Central

Pollution Control Board (CPCB), Government of India (Grant No. AQM/Source Apportionment EPC project/2017), for the funding support to carry out this project.## References

Aiken, A.C., Salcedo, D., Cubison, M.J., Huffman, J.A., DeCarlo, P.F., Ulbrich, I.M., Docherty, K.S., Sueper, D., Kimmel, J.R., Worsnop, D.R., Trimborn, A., Northway, M., Stone, E.A., Schauer, J.J., Volkamer, R.M., Fortner, E., de Foy, B., Wang, J., Laskin, A., Shutthanandan, V., Zheng, J., Zhang, R., Gaffney, J., Marley, N.A., Paredes-Miranda, G., Arnott, W.P., Molina, L.T., Sosa, G., Jimenez, J.L., 2009. Mexico City aerosol analysis during MILAGRO using high resolution aerosol mass spectrometry at the urban supersite (T0) – Part 1: Fine particle composition and organic source apportionment. Atmos. Chem. Phys. 9, 6633–6653. https://doi.org/10.5194/acp-9-6633-2009

Almeida, S.M., Pio, C.A., Freitas, M.C., Reis, M.A., Trancoso, M.A., 2006. Source apportionment of atmospheric urban aerosol based on weekdays/weekend variability: evaluation of road re-suspended dust contribution. Atmos. Environ. 40, 2058–2067. https://doi.org/10.1016/j.atmosenv.2005.11.046

Bao, R., Zhang, A., 2020. Does lockdown reduce air pollution? Evidence from 44 cities in northern China. Sci. Total Environ. 731, 139052. https://doi.org/10.1016/j.scitotenv.2020.139052

Brown, S., Lee, T., Roberts, P., Collett, J., 2016. Wintertime Residential Biomass Burning in Las Vegas, Nevada; Marker Components and Apportionment Methods. Atmosphere (Basel). 7, 58. https://doi.org/10.3390/atmos7040058

Bullock, P., Gregory, P.J., 1991. Soils: A Neglected Resource in Urban Areas, in: Soils in the Urban Environment. Blackwell Publishing Ltd., Oxford, UK, pp. 1–4. https://doi.org/10.1002/9781444310603.ch1

Burr, M.J., Zhang, Y., 2011. Source apportionment of fine particulate matter over the Eastern U.S. Part I: source sensitivity simulations using CMAQ with the Brute Force method. Atmos. Pollut. Res. 2, 300–317. https://doi.org/10.5094/APR.2011.036

Central Electricity Authority, 2012. Report of The Group for Studying Range of Blending of Imported Coal with Domestic Coal. New Delhi.

Chakraborty, A., Mandariya, A.K., Chakraborti, R., Gupta, T., Tripathi, S.N., 2018. Realtime chemical characterization of post monsoon organic aerosols in a polluted urban city: Sources, composition, and
25

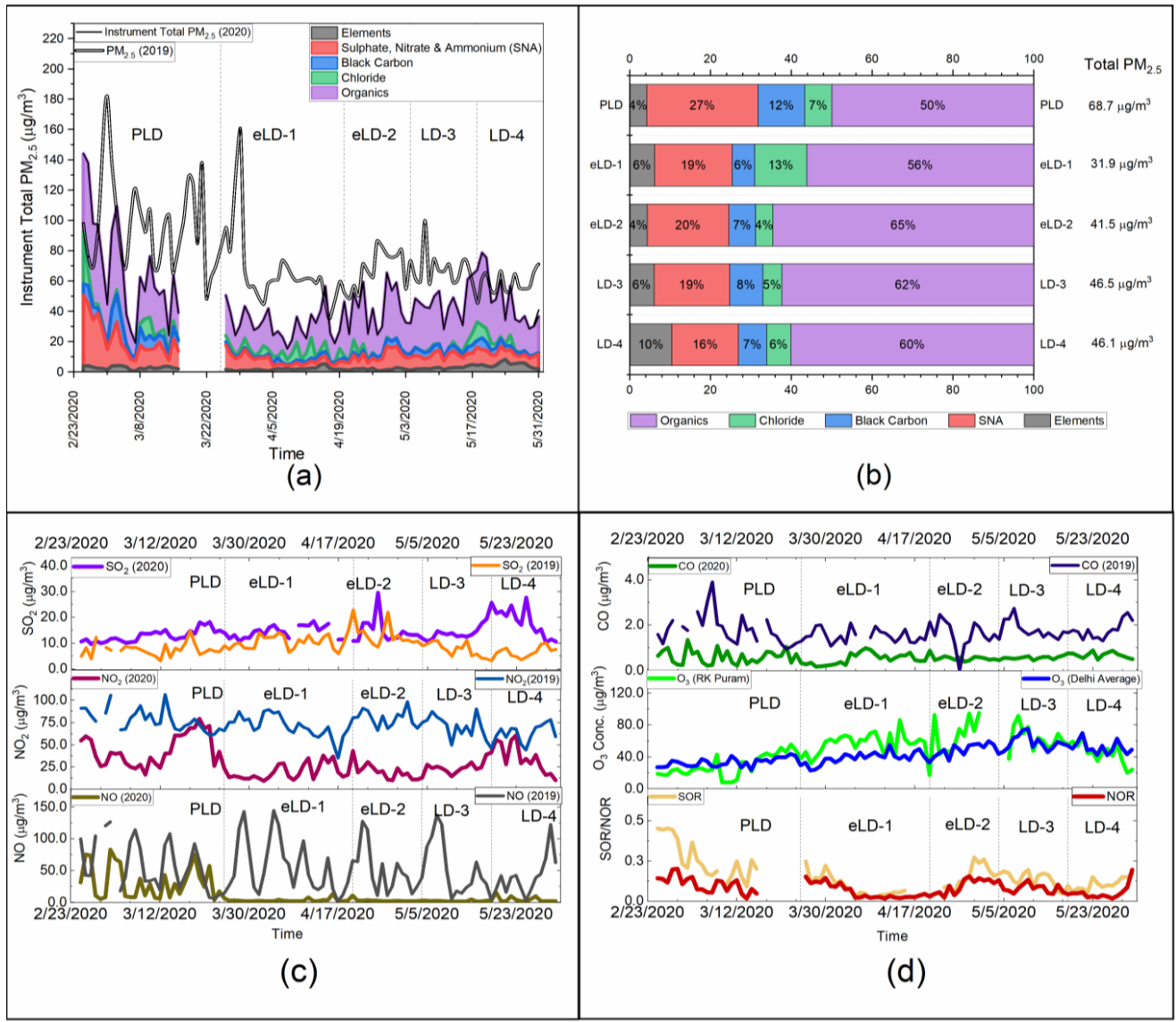

Figure 1: Effect of the lockdown on gaseous pollutants and particulate matter, (a) Time variation of instrument total PM$_{2.5}$ and its major constituents and its comparison with PM$_{2.5}$ levels during corresponding period in 2019 (Instrument Total PM$_{2.5}$ = elemental PM2.5 (Xact) + SNA (ACSM) + Chloride (Xact)+ Black Carbon (Aethalometer)); (b) Phase-wise composition of measured PM$_{2.5}$; (c) Time variation of SO$_2$, NO$_2$, NO (top to bottom) in 2020 and 2019 for corresponding period; (d) Time variation of CO, O$_3$ (at RK Puram and average for Delhi), SOR and NOR (top to bottom) in 2020 and 2019 for corresponding period.

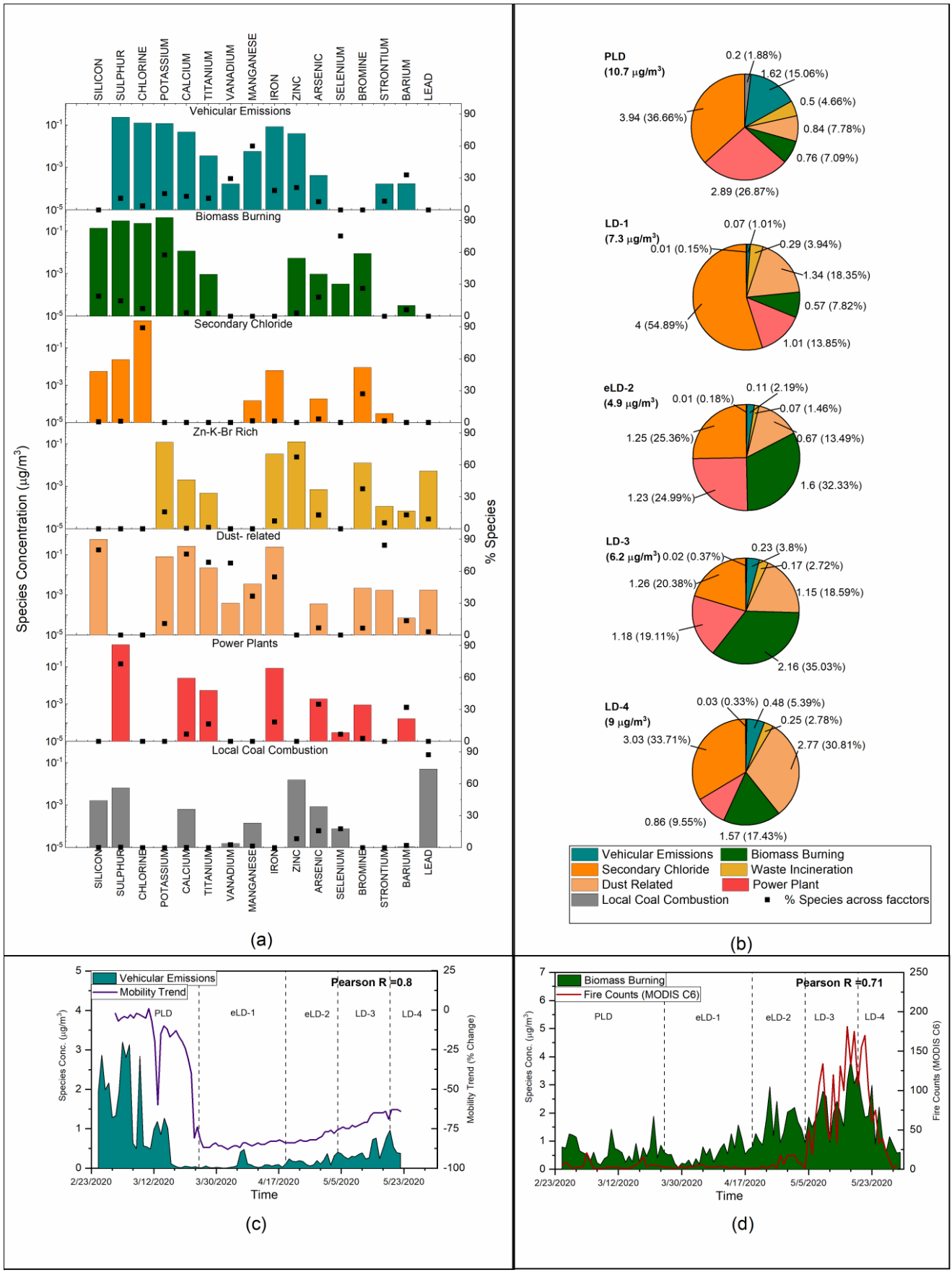

Figure 2: Source apportionment results for elemental particulate matter: (a) resolved source/factor profiles; (b) phase-wise contribution of each factor to total elemental fraction of $PM_{2.5}$; (c) Correlation of vehicular emissions with mobility trends (d) Correlation of biomass burning with fire counts

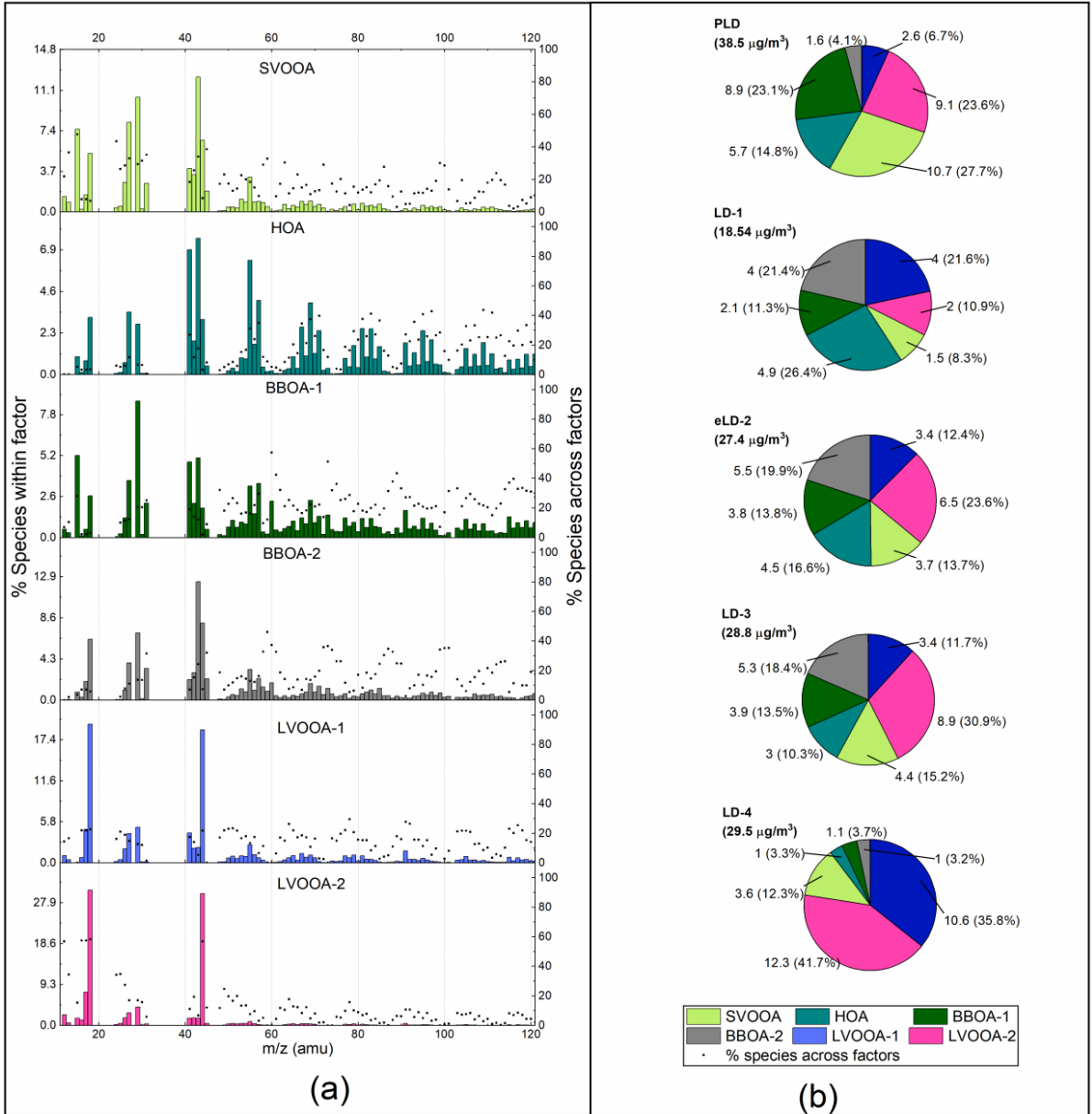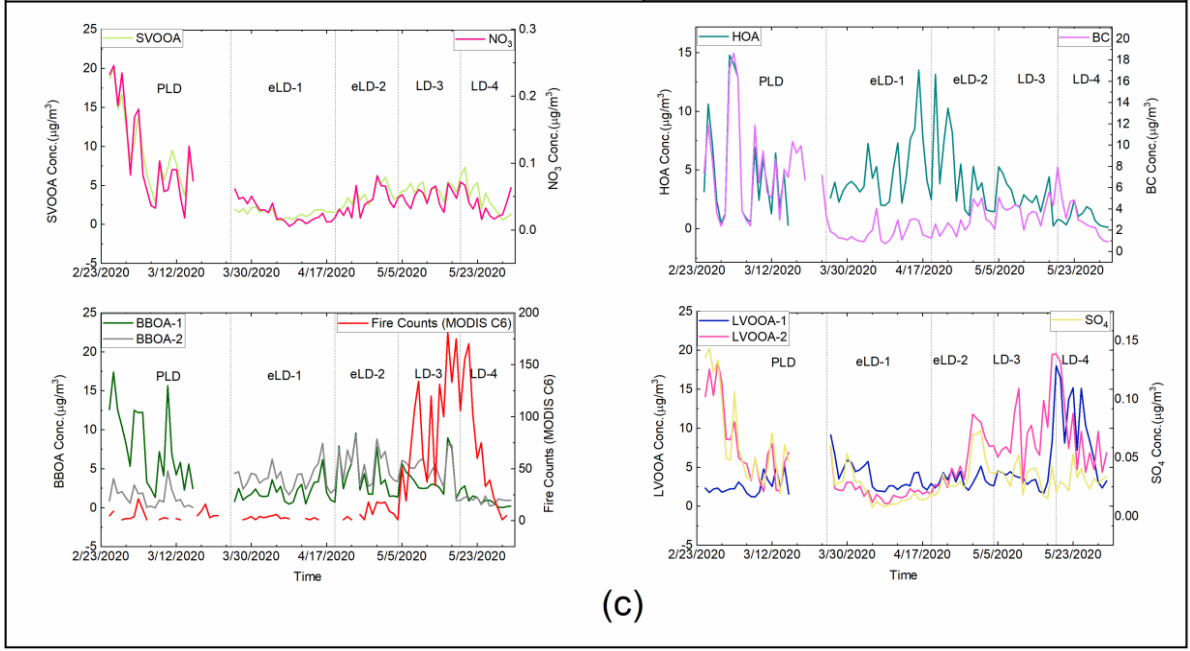

Figure 3: Source apportionment results for organic particulate matter: (a) resolved source/factor profiles; (b) phase-wise contribution of each factor to total organic aerosol; (c) correlation of factors with external markers

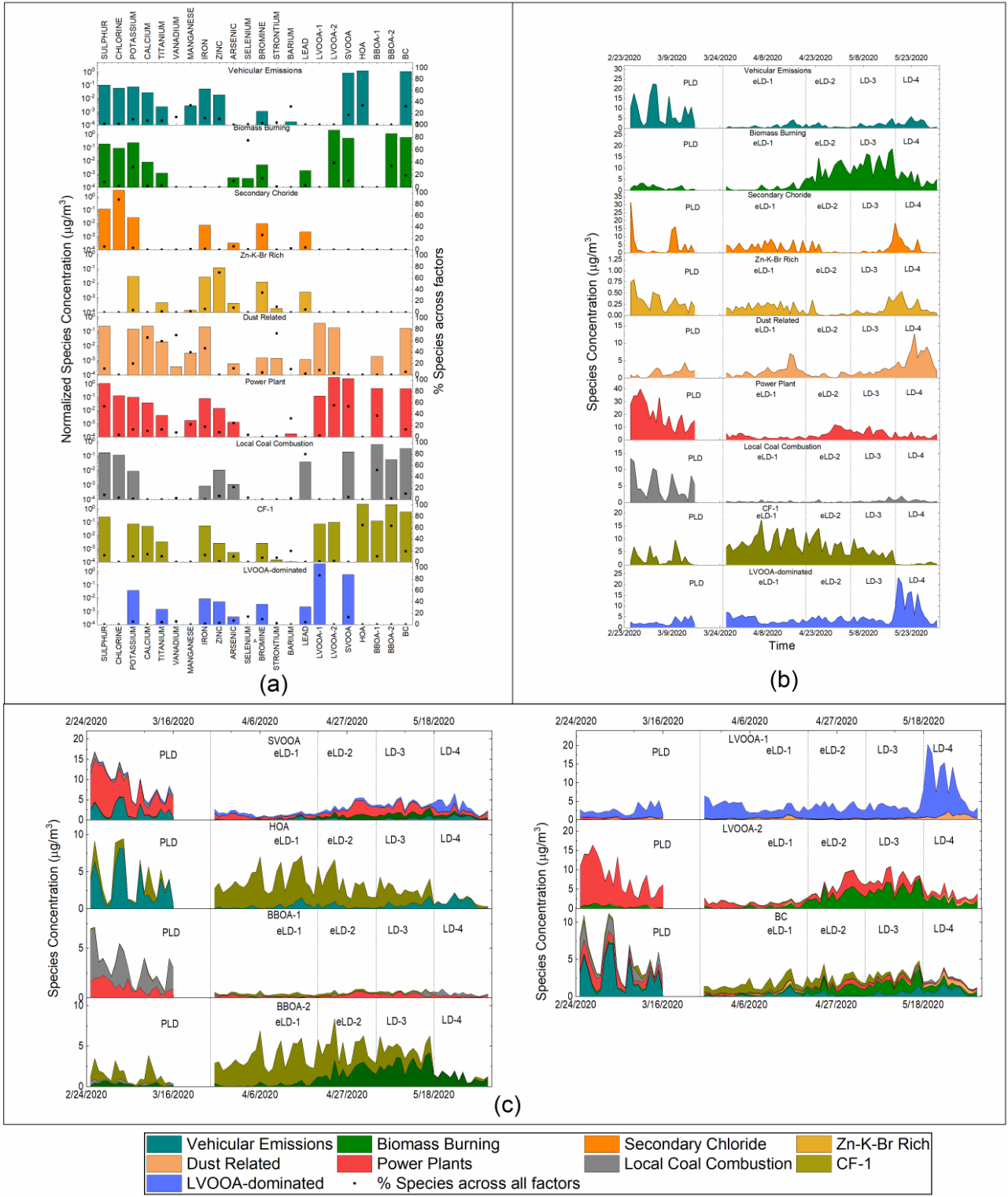

Figure 4: Organic-Inorganic coupled source apportionment: (a) resolved source/factor profiles; (b) temporal variation of each resolved factor; (c) fractional contribution of each double PMF resolved factor to organic SA factors and black carbon

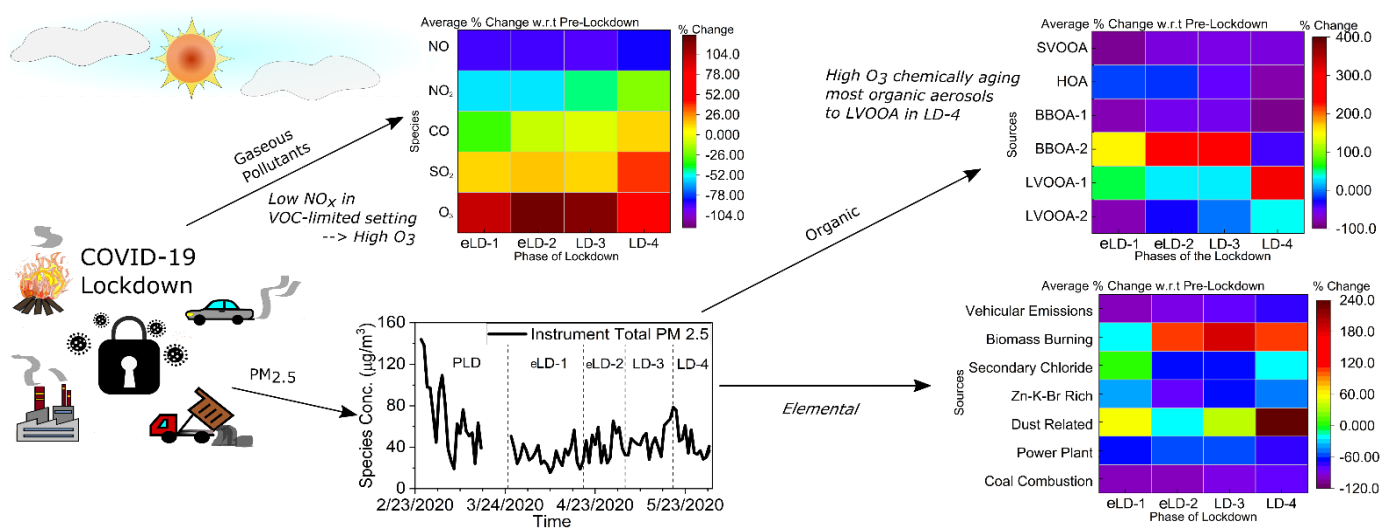

"Graphical Abstract"

# Supplementary Information for "Variation in chemical composition and sources of $PM_{2.5}$ during the COVID-19 lockdown in Delhi."

Chirag Manchanda [a], Mayank Kumar [a,*], Vikram Singh [b,*], Mohd Faisal [b], Naba Hazarika [c], Ashutosh Shukla [d], Vipul Lalchandani [d], Vikas Goel [a], Navaneeth Thamban [d], Dilip Ganguly [e], Sachchida Nand Tripathi [d,*]

a. Department of Mechanical Engineering, Indian Institute of Technology Delhi, New Delhi, India
b. Department of Chemical Engineering, Indian Institute of Technology Delhi, New Delhi, India
c. Department of Applied Mechanics, Indian Institute of Technology Delhi, New Delhi, India
d. Department of Civil Engineering, Indian Institute of Technology Kanpur, Uttar Pradesh, India
e. Centre for Atmospheric Sciences, Indian Institute of Technology Delhi, New Delhi, India

## S1. Quality Assurance and Quality Control (QA/QC) Procedures

The Xact 625i ambient metals monitor was operated at the standard flow rate of 16.7 lpm while sampling ambient air for $PM_{2.5}$ at an hourly time resolution. The instrument is designed to perform automated quality assurance checks every midnight for the elements Cr, Cd, and Pb. In addition to the automated QA checks, leak and flow checks along with XRF and flow calibration was performed at regular intervals as per the maintenance routine advised by the manufacturer.

The Q-ACSM was operated with a standard flow rate of around 0.1 lpm, while sampling and analyzing NR-PM2.5 at a 10-min interval, with background air, allowing for continuous air beam correction, and averaged to an hourly resolution. The ACSM is run using a capture vaporizer, and the corresponding collection efficiency is taken as 1, as suggested by Hu et al., (2017).The ionization efficiency (IE) of $NO_3$ and relative ionization efficiencies (RIE) of $NH_4$ and $SO_4$ were calculated using a monodisperse supply of $NH_4NO_3$ and $(NH_4)_2SO_4$ aerosols, selected through a differential mobility analyzer (DMA) and counted using a condensation particle counter (CPC) as suggested by Crenn et al. (2015). The ACSM calibrations were performed using full scan mode, as per the manufacturer's recommendations (Freney et al., 2019), both before starting the campaign and after its completion. The average

---

[*] Corresponding Authors
E-mail addresses : kmayank@mech.iitd.ac.in (M. Kumar); vs225@chemical@iitd.ac.in (V. Singh), snt@iitk.ac.in (S.N. Tripathi)

Pieber correction i.e. the ratio $CO_2^+$ to total $NO_3$, while calibrating with pure ammonium nitrate calibration was found to be 0.6%; the corresponding median correction to *m/z 44*, was found to be ~5% w.r.t the true *m/z 44* signal, indicating that the impact of the Pieber effect was negligible on the measurements in the present study (Freney et al., 2019).

The Aethalometer was run at a standard operating flow rate of 5 lpm, with a 1-minute time resolution, and averaged to an hourly resolution. While the Aethalometer is designed to operate independently with minimal intervention, the routine was performed as per the manufacturer's recommendation. The routine maintenance involved regular flow rate checks, inspection, and cleaning of the optical chamber and insect screen assembly. The measurements were monitored continuously for any instrument warnings, noise/spikes, as well as BC vs. attenuation values for any required compensation.

The instruments employed in this study (discussed in section 2.2), are designed to measure partial, rather than total $PM_{2.5}$ as none of the three instruments can measure refractory cations and anions like sodium, magnesium, oxides and fluorides. The measurements from these individual instruments were totaled (and referred to as Instrument total $PM_{2.5}$) and regressed against the readings of a collocated BAM measuring total $PM_{2.5}$ during PLD and LD-3 to check for mass closure (during LD-2 and LD-4, some technical issues, which couldn't be addressed adequately during the lockdown, resulted in no sampling). We noted a significant correlation (Pearson R > 0.91) between the total instrument measured particulate mass and the BAM readings, the regression results are presented in figure S1(a) and (b). The daily-average temporal variation of the instrument total $PM_{2.5}$ and the BAM measurements, were compared for the periods when concomitant measurements were available (PLD and LD-3) (Figure S1(c)). The median residual mass (i.e. difference between the BAM measurements and instrument total) was noted to be 10.3%, while the average residual mass was 13.1%, during the campaign.

In addition to comparing the instrument total $PM_{2.5}$ and BAM measurements, we also compare the species common between the two instruments. Chloride is measured by both Xact and ACSM; also the Xact measures sulfur, which can be multiplied by 3 to estimate equivalent sulfate (assuming that all particle-bound sulfur is in form of sulfate, which is also measured by the ACSM. While the aforementioned species are common to both the instruments, Xact measures total sulfur and chlorine, whereas the ACSM can only measure the non-refractory fraction. Figure S7(a) presents the comparison between the Xact-measured and ACSM-measured chloride. We note that from PLD to eLD-3 there is excellent correlation between the two datasets (Pearson R > 0.95) and the time series of both are almost coincident. However, towards the end of LD-3, the Xact measured Cl trends higher as compared to the ACSM measured Cl, while both time series still maintain a good correlation. This behavior can be attributed to the fact that the ACSM only measure the non-refractory fraction of $PM_{2.5}$-bound Cl. It is possible that some refractory chloride salts significantly contribute to the ambient particle-bound chloride during LD-4. Also, during LD-4 there is a rise in biomass burning sources, which may contribute KCl as discussed in section 3.1.2. Past studies have discussed about the deviation of KCl from ideal non-refractory behavior and thus the difference in vaporization kinetics between Cl salts may contribute to the inability of ACSM to account for actual concentration of KCl (Tobler et al., 2020).

Figure S7 (b), demonstrates the variation of Xact derived equivalent sulfate and ACSM sulfate, both signals are observed to maintain an appreciable correlation during each phase of the study (Pearson R> 0 9). However, Xact derived sulfate almost always is greater than or equal to the ACSM sulfate, indicating the difference between the two signals is potentially due to refractory sulfate salts which are measured by the Xact but not by the ACSM.

## S2. Positive Matrix Factorization(PMF) Analysis

### S2.1 PMF Input preparation

The elemental species measured using Xact 625i and the organics fraction measured using the Q-ACSM are subjected to PMF-based source apportionment individually. Xact 625i is set to an hourly time resolution, while the Q-ACSM measures and chemically analyzes the particulate matter every 10-minutes, but the measurements are aggregated to hourly resolution for comparison with the Xact 625i measurements.

Some species (elements for Xact or *m/z* ratios for ACSM) were excluded from the PMF input in cases where an appreciable amount of data points (greater than 50%) are below the Minimum Detection Limit (MDL) (Polissar et al., 1998). The MDLs for each measured elemental species were provided by the manufacturer (Cooper Environmental), while the MDL for the organic fragments was determined according to Singla et al. (2017). for Q-ACSM. Missing data points (due to failure in power supply, routine maintenance) were neglected from the PMF input for this study (Rai et al., 2020b).

The quality of measurement of each species *j* was further characterized based on the signal to noise (S/N) ratio calculated within the EPA PMF 5.0 module (Norris et al., 2014). For every concentration value $x_{ij}$ and corresponding uncertainty $s_{ij}$, the difference between the two is used as the signal $d_{ij}$, such that :

$$d_{ij} = \left(\frac{x_{ij} - s_{ij}}{s_{ij}}\right) \text{ if } x_{ij} > s_{ij} \ldots\ldots (1)$$

$$d_{ij} = 0 \quad \text{if } x_{ij} < s_{ij} \ldots\ldots (2)$$

(S/N) ratio is then given by Equation 3 as follows:

$$\left(\frac{S}{N}\right)_j = \frac{1}{n}\sum_1^n d_{ij} \ldots\ldots (3)$$

The species which had S/N > 1 were categorized as strong in data quality. The input concentration and uncertainty were used for further analysis with no modification. The species with S/N between 0.5 and 1 were categorized as weak, and the input uncertainty values were increased by a factor of four and used along with the input concentration for further analysis. Finally, species with an S/N ratio below 0.5 were classified as bad values and were excluded from further analysis.

Xact 625i was set up to quantify 36 elements (Si, S, Cl, K, Ca, Sc, Ti, V, Mn, Fe, Co, Zn, Ga, Ge, As, Se, Br, Rb, Sr, Mo, Pd, Ag, Cd, In, Sn, Sb, Te, Cs, Ba, La, Ce, Au, Hg, Tl, Pb, and Bi) with an hourly time resolution. During the present study 19 elements (Sc, Co, Ga, Ge, Mo, Pd, Ag, Cd, In, Sn, Sb, Te, Cs, La, Ce, Au, Hg, Tl, Bi) were neglected as more than 50% of the data points were below the MDL for the respective species. Xact 625i software reports measurement uncertainty $s_{ij}$ for every measurement, that accounts for the spectral deconvolution uncertainty as well as the sampling uncertainty (Tremper et al., 2018), the same has been employed for the PMF input in this study

The Q-ACSM employed in our study determines quantitative mass spectra of non-refractory (NR) PM$_{2.5}$ (fraction of total PM$_{2.5}$ that flash vaporizes at 600°C) up to mass to charge ratios (*m/z*) of 200 (Crenn et al., 2015; Ng et al., 2011). However, for the present study, the input was restricted to *m/z 120*, as much lower particle mass was observed for higher *m/z's*. The measured spectrum was resolved into nitrate, sulfate, ammonium, and organic fractions using a library of known fragmentation characteristics (Allan et al., 2004). For the ACSM species, the PMF input uncertainty matrix was calculated using the standard Q-ACSM data analysis software following the procedures described in past studies (Ng et al., 2011; Ulbrich et al., 2009).

*S2.1.1 Double PMF*

As discussed in in the main manuscripts in section 2.2 and 3.3, The present study build upon the work of Petit et al., (2014), which introduced the double PMF methodology initially. For, the first step of the two stage process, the organics only PMF is performed using the Q-ACSM MS and uncertainty, as described above. The steps involved in approaching the PMF solution is discussed upon in section S2.2. As a part of the next step, the factors resolved as a result of the first stage, are used as inputs to a second PMF, along with black carbon data from Aethalometer and elemental data from Xact. As in the case of the elemental PMF, the uncertainty corresponding to the elemental measurements, is taken same as the value uncertainty value reported, by the instrument, also the below MDL values are treated, in the same way as in the case of the elements only PMF. The MDLs for the organic factors and black carbon is adopted from Petit et al., (2014), which has utilized the same instruments as in the present study. The uncertainties and below MDL values for the organic factors and BC was calculated as in a similar fashion as by Petit et al., (2014), using Polissar et al., (1998). The final uncertainty of $j^{th}$ measurement of species $i$ is given by:

$$U_{ij} = \begin{cases} \frac{5}{6} \times MDL_i & if\ C_j < MDL_i \\ \sqrt{u_i^2 \times C_j^2 + MDL_i^2} & if\ C_j > MDL_i \end{cases}$$

Where $u_i$ is the relative uncertainty for species $i$ and $C_j$ is the concentration in the $j^{th}$ row of the input matrix. As documented by Petit et al., (2014), the $u_i$ for the organic factors is derived using the initial organic mass spectra uncertainty. For the organic mass spectra utilized in the present study, the median uncertainty was 17.24%, while mean uncertainty was 21.35%, assuming that the first PMF analysis will add to the uncertainty due to modelling errors (Petit et al., 2014), the relative uncertainty for organic factors was taken as 30% or $u_{OA}$ as 0.3. For BC the relative uncertainty was taken as 30% as an extended uncertainty to the 20% error of BC concentrations due to the Weingartner correction. Finally, to account for variability in the error distribution across the 3 instrument, an additional 10% modelling uncertainty was included in the EPA PMF module, the modelling uncertainty was found to aid in reducing the rotational sensitivity of the solution without impacting the $Q/Q_{exp}$, which is discussed further in section 2.2. Also in line with the recommendations by Petit et al., (2014), each of the both the relative uncertainties corresponding to OA factors and BC concentrations i.e. 30%, were subjected to a sensitivity analysis, by varying one at a time from 30% to 60% at a step size of 10% and evaluating the impact on $Q/Q_{exp}$. However, similar to Petit and co-workers, it is observed that the solution remains unaffected by increasing the uncertainty values and the % change in $Q/Q_{exp}$ is within 2% without a discernable increasing/decreasing trend. This indicates that the solution is governed by the elemental tracers rather than the organic factors or black carbon.

**S2.2 Factor Selection and Uncertainty Quantification**

The selection of the number of factors is an essential user-dependent step in any source apportionment study implemented through PMF. However, since the PMF technique is based on minimization of weighted residual error for the linear fitting of a multivariate system of variables (section S1.1), there are often several solutions possible with similar residual structure, leading multiple past studies to conclude that mathematical diagnostics alone were insufficient for choosing the right number of factors (Canonaco et al., 2013; Rai et al., 2020b). Thus, for the present study the in addition to the diagnostics like total Q-value, $Q/Q_{exp}$ (Canonaco et al., 2013) or scaled residuals for each species, we explore the realizability of each resolved factor by comparing the factor time variation with external tracers, known diurnal trends and known elemental contribution or mass spectra reported in previous studies.

For the initial base runs, we examine the solution space, with the number of factors ranging from 3 to 10 with 10 seeds each (number of PMF runs with different pseudorandom starts) for both ACSM and Xact based source apportionment, along with the second stage of double PMF. In the vicinity (+/-1 factor) of the possible optimal solution (lowest Q-value with no significant change with further increment in the number of factors), the process is repeated with 50 seeds each, similar measures to ensure reproducibility have been adopted by Bhandari et al., (2020). In the present study it is observed that all solutions in the vicinity of the aforementioned optimal solution, are highly reproducible when the process is repeated over 50 seed runs, with less than 1% change in min Q-value.

After initial screening of the solution with the Q-value or $Q/Q_{exp,}$ the optimal solution (in all three PMF analysis undertaken in the present study) is tested for rotational ambiguity by subjecting it to DISP analysis available within the EPA PMF 5.0 module. The DISP assess the largest range of solution factor profile values without an appreciable change in the optimal Q-value. In the DISP method, each species in the factor profiles obtained from the base-run are perturbed about the base value, one species at a time and after each adjustment the PMF run is repeated to calculate a new converged solution such that the change in the Q-value, w.r.t the base run remains less than a predetermined maximum change i.e. $dQ_{max}$ ($dQ_{max}$ = 4,8,15,25). With these changes in the factor profiles, it is possible that the modified factor profile may switch identity when compared to the base run factors i.e. a particular factor from the base run after displacement of certain species may be better correlated to some other base factor than itself, such a case is noted as a factor swap. The EPA PMF module accounts for factor swaps using uncenterd cross correlations between the displaced solution and the base run. In the present study no factor swaps were reported for both the Xact-based and ACSM-based PMF solutions. The largest decrease in Q-value for the Xact-based PMF was found to be 0.0668 with % dQ change as 0.002%. For the ACSM-based source apportionment the largest decrease in Q-value was observed to be 0.2028 with % dQ change as 0.0032%. For the second stage of the double PMF the largest decrease in Q-value was observed to be 0.6946 with % dQ change as 0.017%. As per Norris et al. (2014) % dQ change under 1% represents an acceptable DISP solution, representing reliable PMF solution.

However, the optimal base run solutions was further subjected to variation in f-peak values from -1 to +1 at an increment of 0.1 to explore the rotations of plausible solutions, to evaluate the effect of rotations on the fraction of variance explained by each factor, correlation of factors' mass spectra (MS) with reference mass spectra (for ACSM), correlation of factor time series with external tracers and the mutual correlation between the time series of resolved factors (Bhandari et al., 2020; Rai et al., 2020b; Ulbrich et al., 2009). The range of f-peak rotations was limited to ensure that the rotated solution remains in the vicinity (similar total Q-value) of the initial solution (except the case of a rotationally ambiguous solution) in the solution space.

For the Xact-based source apportionment, as well as the double PMF, no appreciable change in the factor source profiles, and external correlation was observed by varying f-peak; thus, the base solution was chosen as optimal. However, for the ACSM-based source, apportionment rotations were found to improve correlations between factor MS of all factors and reference MS from Ng et al. (2011) and also decrease mutual correlations between the resolved factors. Thus the rotated solution with the best correlations with reference MS and minimal mutual correlations among factors was chosen to be optimal (Bhandari et al., 2020; Rai et al., 2020b).

As discussed in section S2.1.1, in case of the second stage of the double PMF, an additional modelling uncertainty was introduced in order to harmonize any potential impact from difference in error structures of the 3 instruments employed in the present study (Petit et al., 2014). Modelling uncertainty was varied from 0% to 20% in steps of 5% each and its impact its impact on sensitivity to rotation by f-peak was analyzed. It is observed that introduction of modelling uncertainty reduced the sensitivity to f-peak rotations, % dQ for f = + 0.1 was noted to be 0.25, 0.13 and 0.097 at 0%, 5% and 10% extra modelling uncertainty, however, any further increase in modelling uncertainty had negligible impact on the sensitivity to rotations, thus a 10% modelling uncertainty was deemed optimal (Ulbrich et al., 2009).

Further, the effect of random errors were evaluated using the bootstrap (BS) randomized resampling strategy (Brown et al., 2015; Efron, 1979). The bootstrap is implemented using US EPA PMF 5.0 by a random selection of non-overlapping blocks of species measurements from the input data set and creates a new PMF input matrix with the total number of samples equal to the original input matrix, where the user specifies the block size. The PMF code then runs over the new input matrix, and the BS factors are then mapped to the primary factors. The BS factors are assigned to corresponding base factors with which have the highest uncentered correlation values (above a user-specified threshold). If a particular BS factor doesn't have an uncentered-correlation higher than the threshold with any base factor, then it is considered to be "unmapped." This analysis provides us with a proxy to understand the uncertainty associated with the solution and the apportionment of each species in the resolved factors.

In the present study, the solutions deemed optimal following the prescreening based on Q/Qexp and further evaluation based on f-peak rotations and DISP analysis, were subjected to 800 BS runs each, with a threshold correlation of 0.8. In the case of the organics-only SA 764 BS runs were classified as good solutions, having no unmapped factors, while the elemental SA and the double PMF derived solutions resulted in 789 and 782 good solutions respectively (Brown et al., 2015; Norris et al., 2014).

Finally, each solution tested for statistical optimality based on the procedures documented above were tested for environmental realizability. For the elemental SA each factor profile was tested for realizability of based on each element apportioned to the factor, also the temporal variation of the resolved source is correlated to available external indicators/tracers the same has been discussed in detail in section 3.1. In case of the organic only source apportionment, the resolved factors are tested for correlation to established source profiles (Ng et al., 2011), while their temporal evolution is checked for correlation to known external markers like black carbon, sulfate and nitrate, the ame has been discussed in detail in section 3.2. In addition to this both the elemental SA and organic SA derived factors are checked for realizability based on their diurnal variability, which is discussed in further detail in section S3.2.

**S3. Supplementary Results**

The COVID-induced lockdown in India, lasted over a period of 70 days starting 25th March 2020 up to 31st May 2020. The lockdown progressed in a phased manner, during the first phase of the lockdown (LD-1) starting 25th March un till 14th April marked the strictest phase of the lockdown, with nearly all services and commercial activities completely suspended, with an exception to providers of essential goods and services like hospitals, grocery stores, and pharmacies. The lockdown was extended with phase-2 (LD-2) lasting up to 3rd April 2020. However, the first set of relaxations were implemented starting 20th April 2020 with allowances to agricultural industries, farming supplies, cargo services. In order to capture the impact of these relaxations in the present study, we focus on eLD-1 and eLD-2, as defined in section 2.1, instead of LD-1 and LD-2. With the start of LD-3 from 4th May 2020, most regions were subdivided into green, orange, and red-zones based on the intensity of the spread of the virus in the region. In green zones, normal movement was restored with public buses running at 50% capacity, only movement with private vehicles was allowed in orange zones, and no relaxations in red zones. LD-4 started on 18th May with further increase in relaxations, the interstate movement was permitted, all categories of small-scale shops were allowed to open, and all industries allowed to restart, private offices were allowed to reopen with 33% staff.

The following sections draw further support to the main manuscript, evaluating the impact of the aforementioned phased relaxations in the lockdown on the variation of sources contributing to ambient $PM_{2.5}$:

**S3.1 Time variation of different aerosols during lockdown**

*S3.1.1 $PM_{2.5}$ and its constituents*

Analyzing the daily average time variation of the instrument total $PM_{2.5}$ and the fractional contribution of each constituent (Figure 1(a)), as discussed earlier in section 3, we note that the average $PM_{2.5}$ values fall by 53.6% from pre-lockdown to lockdown phase-1. We can also observe that each of the $PM_{2.5}$ constituents, i.e., elements, sulfate, nitrate and ammonium (SNA), black carbon (BC), and organics, is found to decrease by 31%, 67.7%, 77.2%, and 47.8% respectively, between PLD and lockdown eLD-1 (Figure 1(b)). However, chloride remains almost unaffected or minimally affected, with only a 12% decrease due to the lockdown.

From Figure 1 (a) and Figure 1 (b), we can also note that total $PM_{2.5}$ along with the elements, SNA, BC, and organics steadily trend back towards their initial concentrations with increasing relaxations in subsequent phases of the lockdown. However, even in the final phase of the lockdown, total $PM_{2.5}$ was 33% lower when compared to the PLD values, while SNA, BC, and organics remained 59.8%, 59.5%, and 19.2% lower than their PLD concentrations.

While average chloride concentration was 46% lower in LD-4 as compared to the pre-lockdown phase, it seems to be affected more by metrological conditions rather than the lockdown. Firstly, chloride concentration was only marginally affected in eLD-1; also, it reached its minimum average concentration (1.8 $\mu g/m^3$) in eLD-2 rather than eLD-1(Figure 1 (b)). This coincides with the observation that a significant portion of eLD-2 and beginning of LD-3 is affected by disturbances from the southeast (from the wind direction/wind speed, Figure S2(a)). This dependence of chloride on wind direction or other metrological conditions is concomitant with the observations made for the apportioned secondary chloride factor in section 3.1.3. Also, it is interesting to note that the elemental concentration increased by 65% in LD-4 w.r.t the PLD concentrations. This could possibly be due to the increase in gust/dust storm events during LD-4 leading to the rise in apportioned dust factor, as discussed in section 3.1.5.

*S3.1.2 Gaseous pollutants*

In Figure 1 (c) & (d) we plot the concentrations of gaseous pollutants ($SO_2$, $NO_2$, NO, and CO) taken from the Continuous Ambient Air Quality Monitoring Station (CAAQMS) at RK Puram. As discussed in section 3 and presented in Figure S6(a), a significant drop in $NO_2$ and NO concentrations 56% and 90%, respectively, after the lockdown, while $NO_2$ seems to rise back with increasing relaxations, NO concentrations do not rise at the same pace. During LD-4, the average $NO_2$ levels were 22% lower than its pre-lockdown concentration, while for NO, the average concentration was 85% lower than its PLD average. Transportation is believed to be the primary source of both NO and $NO_2$ emissions in Delhi (Tyagi et al., 2016). Similar anomalies have been reported by Nagpure et al. (2013) and Badarinath et al. (2009) in Delhi. However, the disparate behavior could stem from chain reactions initiated by the attack of hydroxyl radical on VOCs and CO catalyzing the conversion of NO to $NO_2$ (Tiwari et al., 2015), outside the photo stationary state of reactions between NO, $NO_2$, and $O_3$ as described by Leighton (1961).

Similarly, CO also follows the intuitive trend of a sudden reduction (by 32%) in LD-4, followed by a gradual increase with increasing relaxations in the lockdown. It is again interesting to note that $SO_2$ remains mostly unaffected by the lockdown based on measurements for the RK Puram station; this is in line with a recent study by Kumari and Toshniwal (2020), where only a slight decrease in $SO_2$ was reported for Delhi, Kumari and co-workers supported this observation with the hypothesis that $SO_2$ in Delhi mainly stems from power plant emissions which have remained unaffected by the lockdown. However, the results of the present study and an independent report from the Power System Operation Corporation (POSOCO, 2020) indicate that the contribution of the power plant emissions to particulate concentrations was indeed impacted by the lockdown, as discussed in section 3.1.6. The potential cause to the limited impact of the lockdown on $SO_2$ levels maybe that $SO_2$ levels are supplemented by some other emission source instead of the power plant source during the lockdown, similar to the HOA supplemented by CF-1 instead of vehicular emissions during the lockdown as discussed in section 3.3. It is also important to note that while the $SO_2$ levels are not strongly impacted by the lockdown, also highlighted by the low SOR levels discussed in section 3 of the main manuscript, the overall particulate bound sulfur/sulfate levels indeed drop significantly during the lockdown. A recent study by Wang et al., (2020) analyzing the driving factors of sulfate formation in $PM_{2.5}$, highlights the role of oxidants like $O_3$ and $NO_x$ in stimulating the conversion from $SO_2$ to sulfate which in turn condenses to the particulate matter. Thus, the sudden noticeable reduction in $NO_x$ levels in eLD-1, as discussed in section 3, could be a potential reason for the disparity between the impact of the lockdown on sulfur in the gaseous and particulate phases.

In Figure S2(b), we further analyze the effect of the lockdown on ambient air quality based on aerosol neutralization ratio (ANR), as defined by Zhang et al. (2007) The ANR was found to be weakly affected by the lockdown for the PLD phase; the average ANR was 0.83, followed by 0.73, 0.77, 0.78, 0.78 for each of eLD-1 to LD-4. The ANR indicated weak acidity associated with the aerosols, which is slightly intensified by the lockdown, followed by a gradual trend back towards the PLD value.

*S3.1.3 Markers for organic sources*

While section 3.2 elaborates on the source apportionment of the organic fraction of the particulate matter, in Figure S2(b), we plot some significant *m/z* ratios as tracers to known apportioned factors like *m/z 43* for semi-volatile OOA (SVOOA) and *m/z44* for low volatility OOA (LVOOA), *m/z 60 & m/z 73* for Biomass Burning Organic Aerosol (BBOA) and *m/z 55 & m/z 57* for Hydrocarbon like Organic Aerosol (HOA). Interestingly, we observe that most of these *m/z's* decrease significantly during lockdown phase-4, while *m/z 44* increases significantly for the same period. Usually, *m/z 55* and *m/z 57* and the apportioned factor marked by them, i.e., HOA, is found to correlate well with vehicular markers like BC (DeWitt et al., 2015), however, in the present case, the reduction in gaseous pollutants like $NO_2$ and NO, also associated with vehicular origin (Tyagi et al., 2016), due to the lockdown is much more intense than observed in *m/z 55* or *57*, supporting possible alternate sources for the same during the lockdown as discussed in section 3.2 with the source apportionment of the organic $PM_{2.5}$.

**S3.2 Diurnal Variation of Factors apportioned using PMF**

The details on the factors resolved from source apportionment of Xact and ACSM derived species are discussed in section 3.1 and 3.2, respectively. In this section, we present the diurnality associated with each resolved factor and the implications associated with the same.

Figure S4 displays the diurnal variation of each factor constituting elemental $PM_{2.5}$, for each phase from PLD to LD-4. The vehicular emissions factor shows appreciable diurnal variation during PLD and LD-4, with significant peaks around 6:00 IST to 8:00 IST and also a relatively weaker peak around 18:00 IST to 21:00 IST, both of which correspond to traffic rush hours in the vicinity of the sampling location; also some peaks after midnight may signal toward the heavy-duty vehicular movement (for logistics and transportation) during the night. However, it is essential to note that the lockdown severely impacts this typical diurnality associated with this factor, and no characteristic peaks are observed during eLD-1 to LD-4, it is essential to consider that the average factor concentration was noted to drop by 96% as discussed in section 3.1. Also, the restricted vehicular movement during the lockdown is not expected to follow a significant trend, justifying the atypical nature of the diurnal pattern.

The biomass burning factor is found to peak in the early morning around 6:00 IST, indicating the influence of gas-particle partitioning and ambient temperature on the time variation of this factor, which is in line with the findings of past studies (Rai et al., 2020a). Also, the nature of the diurnal profile has remained very similar during all phases of the lockdown, indicating a less pronounced effect of the lockdown on the nature of the source.

The diurnality associated with the secondary chloride factor is discussed in section 3.1.3 and is indicative of dependence of the concentration on gas-particle due to increment in volatility and consequent evaporation after sunrise supporting the significant peak around 6:00 IST followed by the rapid decline in concentration. The Zn-K-Br rich factor is found to have diurnal behavior similar to secondary chloride and could partially contribute to the same as a potential source of HCl and HBr, as discussed in section 3.1.3 and 3.1.4. However, it is vital to take into account that secondary chloride remained mostly unaffected by the lockdown, while the Zn-K-Br rich source was found to drop by 42% in eLD-1.

In the present study, while the lockdown is found to have little or no impact on the dust-related factor in terms of the concentration, however, there is significant variation in the diurnal pattern of the same across different phases of the lockdown. But since this factor is expected to be influenced by a variety of metrological and transport problems, it is difficult to attribute the variation to a single cause.

The power plants factor is found to have a relatively weak diurnal variation; the S-rich factor was noted to correlate well with SOR in section 3.1.5, indicating the factor to be driven by sulfate emissions, the weak diurnality can be attributed to the low volatility associated with sulfate; also the peaks around 8:00 IST indicate the rise in factor concentration due to photo-oxidation from $SO_2$ to $SO_4^{2-}$, along with gas-particle partitioning of compounds like $(NH_4)_2SO_4$.

The coal combustion diurnal profile displays a peak around 6:00 IST, similar to secondary chloride signaling towards the role of gas-particle partitioning in giving rise to the observed diurnality.

The diurnal variation profiles of the factors resolved from the source apportionments of organic fraction of NR-PM$_{2.5}$ are collated in Figure S5. The diurnal variation of SVOOA has been discussed in section 3.2.1 and is found to be dependent on photo-oxidation and boundary layer height. The diurnal profile of HOA is typical of a primary organic aerosol with seeming high dependence on boundary layer height, leading to a rise in concentration when the boundary layer height is lower, i.e., during the night and following a decrease in the morning when the boundary layer rises. Overall the diurnal profile has no discernable rush hour peaks to attribute the factor to have a vehicular origin.

Similar to HOA, both BBOA-1 and BBOA-2 display diurnal behavior characteristic of primary organic emissions and their dependence on boundary layer height or ambient temperature (Bhandari et al., 2020), however BBOA-2 in addition to the increased concentration during the night, also presents a distinct peak around 6:00 IST - 9:00 IST, similar to biomass burning factor from Xact source apportionment, indicating the role of gas-particle partitioning or a primary emission associated with that time.

A weak diurnal variation is observed in case of LVOOA-2, the factor displays a nominal peak around noon, which may be attributed to enhanced photo-oxidation of fresh organic emissions, resulting in the formation of oxidized organic aerosols (Aiken et al., 2009; Zhu et al., 2018). However, as discussed in section 3.2.4, LVOOA-1 display diurnal behavior similar to primary organic aerosols, and this behavior is further intensified during LD-4 when the highest average concentration corresponding to this factor is recorded. This anomalous behavior indicates the local formation/emission of LVOOA rather than the typical aging route citing the atypical diurnal as well as other factors discussed in section 3.2.4

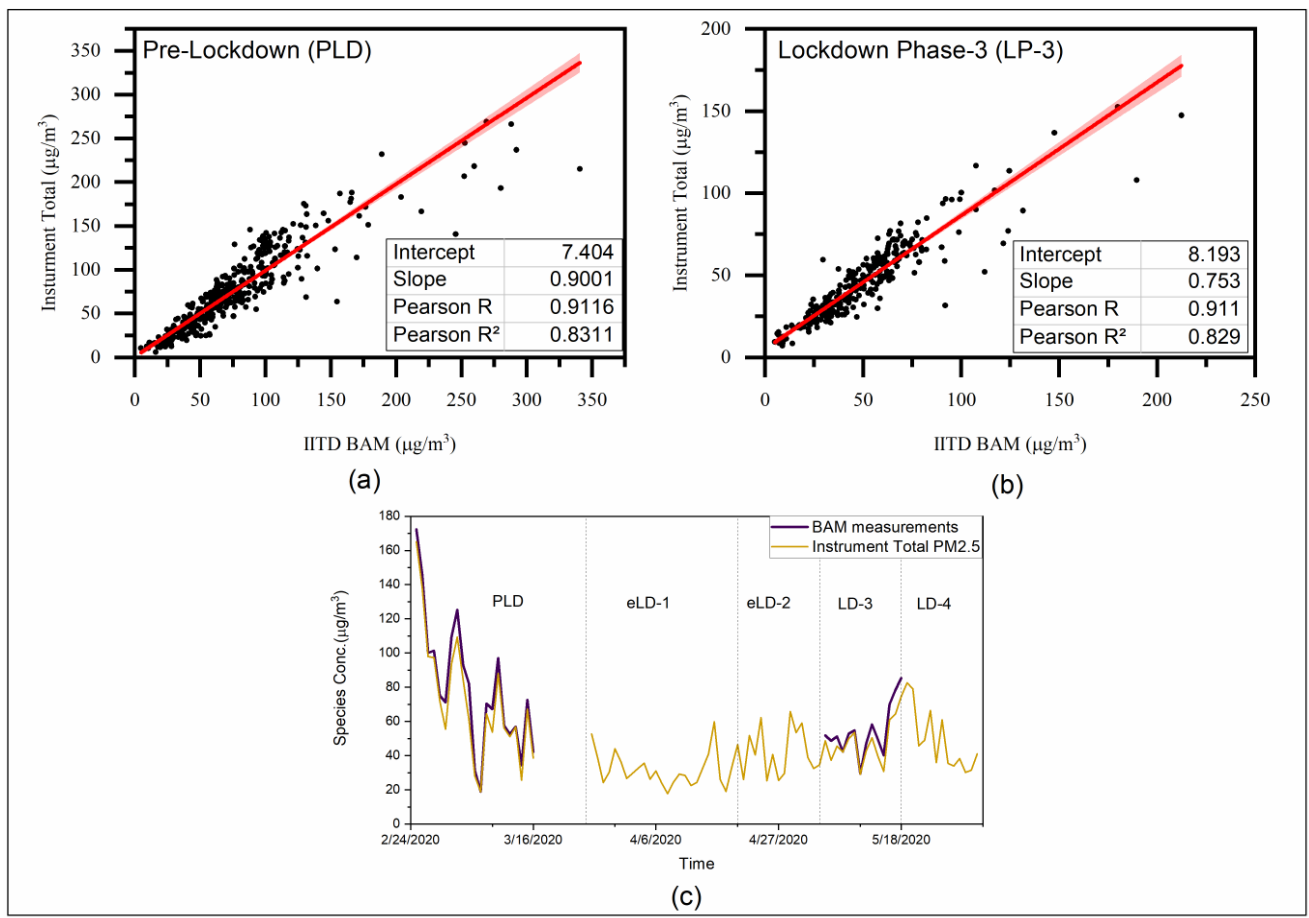

Figure S1: Linear regression analysis between sum of total $PM_{2.5}$ as measured by Xact, ACSM and Aethalometer and a co-located Beta Attenuation Monitor (BAM)

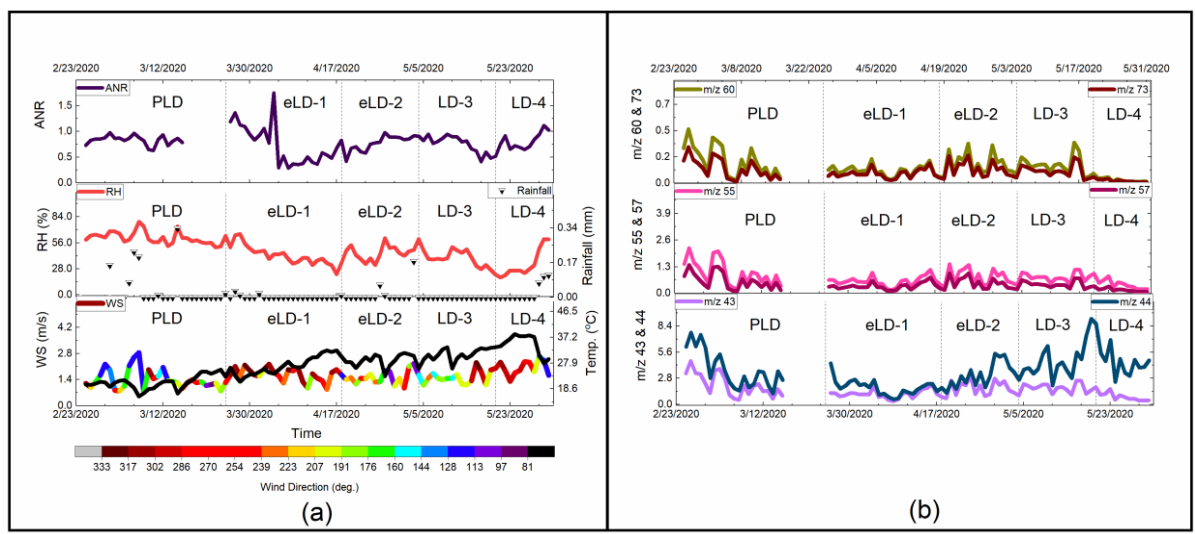

Figure S2: (a) Time variation of ANR, RH WS, WD and ambient temperature through different phases of the lockdown; (b) Time variation of characteristic m/z ratios i.e. 60, 73, 55, 57, 43 and 44 amu, through different phases of the lockdown

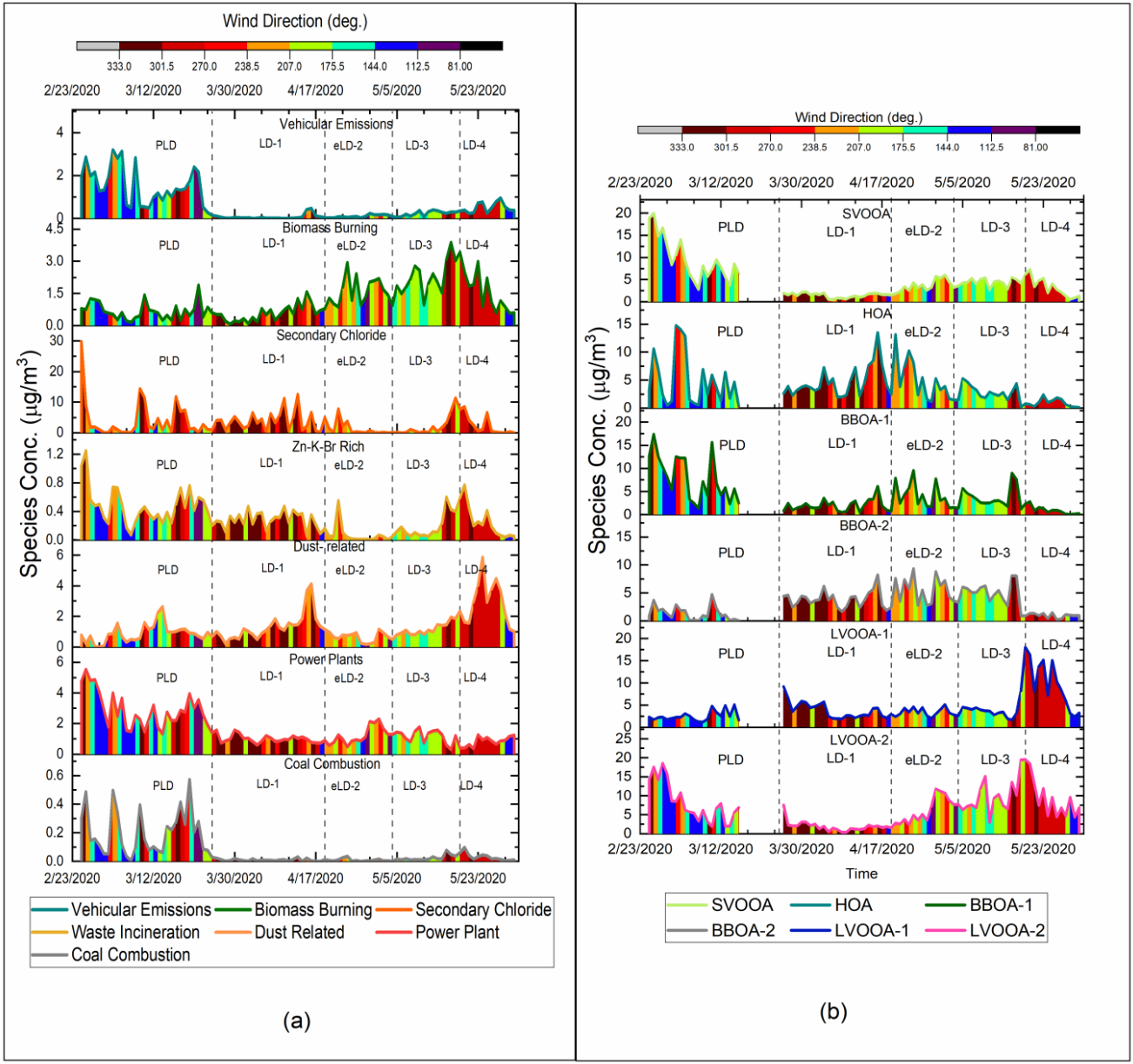

Figure S3: (a) Temporal variation of sources contributing to elemental $PM_{2.5}$ through different lockdown phases (b) Temporal variation of sources contributing to organic $PM_{2.5}$ during different lockdown phases

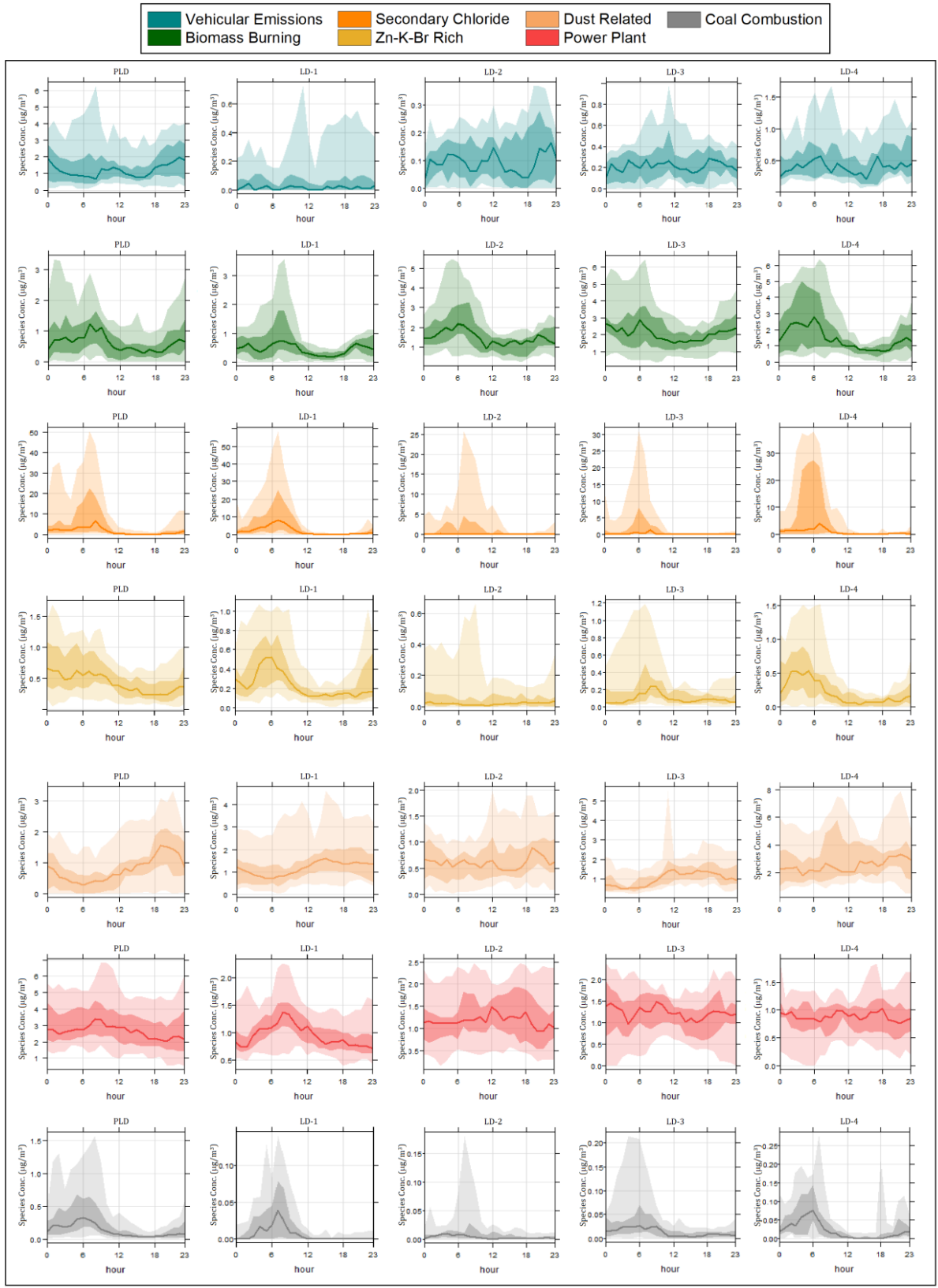

Figure S4: Diurnal Variation of factors resolved from source apportionment of elemental PM$_{2.5}$ (median, 25/75$^{th}$ & 5/95$^{th}$ quantiles)

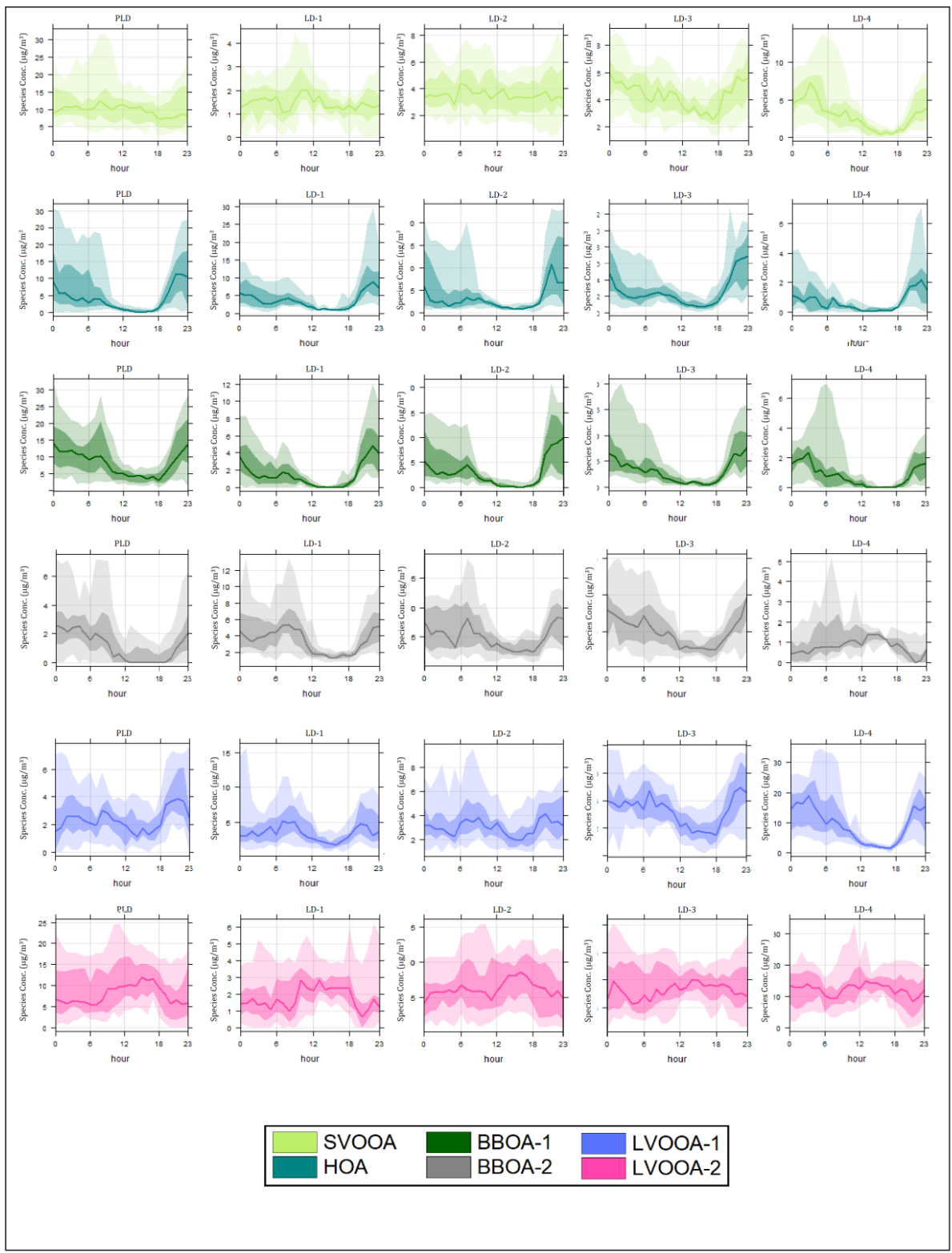

Figure S5: Diurnal Variation of factors resolved from source apportionment of organic NR-PM$_{2.5}$ (median, 25/75$^{th}$ & 5/95$^{th}$ quantiles)

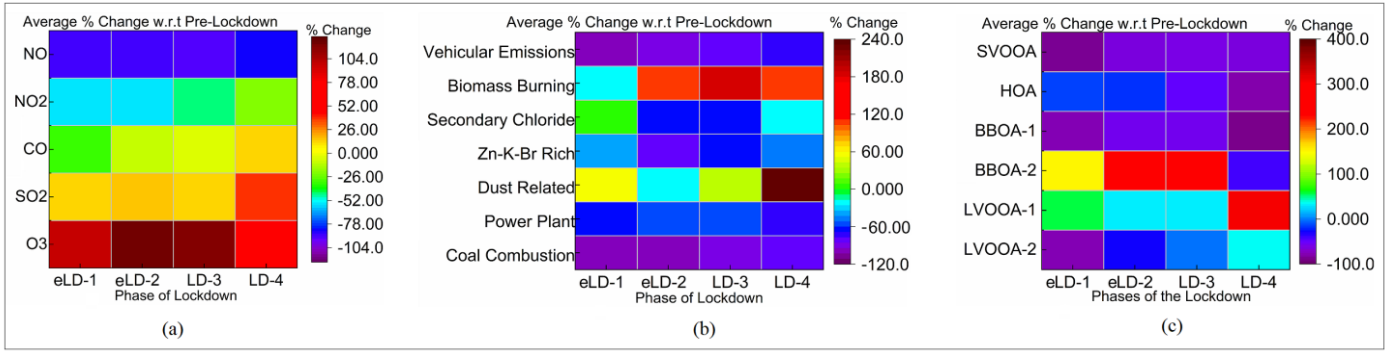

Figure S6: (a) % Change in average gaseous concentrations through each phase of the lockdown w.r.t. pre-lockdown levels; (b) % Change in average concentration of sources contributing to elemental $PM_{2.5}$ through each phase of the lockdown w.r.t. pre-lockdown levels; (c) % Change in average concentration of sources contributing to organic $PM_{2.5}$ through each phase of the lockdown w.r.t. pre-lockdown levels

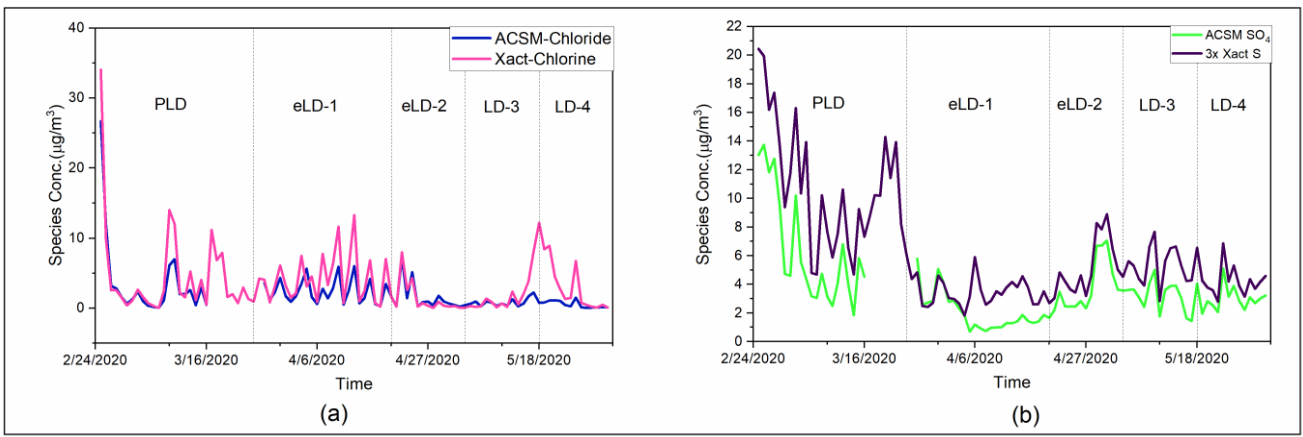

Figure S7: (a) Comparison between ambient chlorine levels measured using the Xact and ACSM during the study period; (b) Comparison between ambient sulfate levels measured using the Xact and ACSM (Sulfur concentrations measured using the Xact are multiplied by 3 to derive equivalent sulfate assuming all particle-bound sulfur exists in form of sulfate).